\documentstyle[12pt]{article}
\def\baselinestretch{1.2}
\input epsf
\catcode`\@=11
\def\grafix#1,#2,#3,#4,#5{\vbadness 10000
    \setbox1\vbox{\vbox to #1{\hbox to #2{}}
                          \vbadness 1000}
    \special{*#3,#4,#5}
    \centerline{\vbox{\hrule
                \hbox{\vrule\hskip 1pt
                       \vbox{\vskip 1pt
                             \box1
                             \vskip 1pt}
                       \hskip 1pt\vrule}
                      \hrule}}}
\@addtoreset{equation}{section}
\def\ksection{\arabic{section}}

\def\@normalsize{\@setsize\normalsize{15pt}\xiipt\@xiipt
\abovedisplayskip 14pt plus3pt minus3pt%
\belowdisplayskip \abovedisplayskip
\abovedisplayshortskip  \z@ plus3pt%
\belowdisplayshortskip  7pt plus3.5pt minus0pt}

\def\small{\@setsize\small{13.6pt}\xipt\@xipt
\abovedisplayskip 16pt plus3pt minus3pt%
\belowdisplayskip \abovedisplayskip
\abovedisplayshortskip  \z@ plus3pt%
\belowdisplayshortskip  7pt plus3.5pt minus0pt
\def\@listi{\parsep 4.5pt plus 2pt minus 1pt
            \itemsep \parsep
            \topsep 9pt plus 3pt minus 3pt}}

\def\underline#1{\relax\ifmmode\@@underline#1\else
        $\@@underline{\hbox{#1}}$\relax\fi}
\@twosidetrue

\catcode`@=12

\catcode`\@=11

\def\thesection{\Roman{section}.}

\def\FERMIPUB{}
\def\FERMILABPub#1{\def\FERMIPUB{#1}}
\def\ps@headings{\def\@oddfoot{}\def\@evenfoot{}
\def\@oddhead{\hbox{}\hfill
        \makebox[.5\textwidth]{\raggedright\ignorespaces
--\thepage{}--
        \hfill {\rm FERMILAB--Pub--\FERMIPUB}}}
\def\@evenhead{\@oddhead}
\def\subsectionmark##1{\markboth{##1}{}}
}

\def\a{\alpha}
\def\b{\beta}

\def\d{\delta}
\def\D{\Delta}
\def\e{\epsilon}

\def\F{\Phi}

\def\g{\gamma}
\def\G{\Gamma}
\def\h{\eta}

\def\k{\kappa}
\def\l{\lambda}
\def\L{\Lambda}
\def\m{\mu}
\def\n{\nu}

\def\p{\pi}

\def\q{\theta}

\def\s{\sigma}

\def\t{\tau}

\def\U{\Upsilon}

\def\fr{\frac}
\def\ba{\begin{array}}
\def\ea{\end{array}}
\def\bz{\begin{equation}}
\def\ez{\end{equation}}
\def\by{\begin{eqnarray}}
\def\ey{\end{eqnarray}}

\def\nn{\nonumber}
\newcommand{\ls}[1]
   {\dimen0=\fontdimen6\the\font
    \lineskip=#1\dimen0
    \advance\lineskip.5\fontdimen5\the\font
    \advance\lineskip-\dimen0
    \lineskiplimit=.9\lineskip
    \baselineskip=\lineskip
    \advance\baselineskip\dimen0
    \normallineskip\lineskip
    \normallineskiplimit\lineskiplimit
    \normalbaselineskip\baselineskip
    \ignorespaces}

\def\ctp{\tilde{\p}^{\pm}}
\def\ch2{\tilde{H}^{\pm}}

\ps@headings

\catcode`\@=12

\relax

\newcounter{appendix}

\def\appendix{\par
 \addtocounter{appendix}{1}
 \def\thesection{Appendix \Alph{appendix}:}
 \def\ksection{\Alph{appendix}}}

\newskip\humongous \humongous=0pt plus 1000pt minus 1000pt

\newif\ifdtup

\def\oldreffmt#1{\rlap{[#1]} \hbox to 2\parindent{}}

\def\figfmt#1{\rlap{Figure {#1}} \hbox to 1in{}}

\def\VEV#1{\left\langle #1\right\rangle}

\def\beq{\begin{equation}}
\def\eeq{\end{equation}}
\def\bea{\begin{eqnarray}}
\def\eea{\end{eqnarray}}
\def\half{\frac{1}{2}}

\def\bq{\begin{quote}}
\def\eq{\end{quote}}

\def\half{\frac{1}{2}}
\def \gta {\mathrel{\vcenter
     {\hbox{$>$}\nointerlineskip\hbox{$\sim$}}}}
\newcommand{\be}{\begin{equation}}
\newcommand{\ee}{\end{equation}}
\newcommand{\bdm}{\begin{displaymath}}
\newcommand{\edm}{\end{displaymath}}
\def\simlt{\mathrel{\lower2.5pt\vbox{\lineskip=0pt\baselineskip=0pt
           \hbox{$<$}\hbox{$\sim$}}}}
\def\simgt{\mathrel{\lower2.5pt\vbox{\lineskip=0pt\baselineskip=0pt
           \hbox{$>$}\hbox{$\sim$}}}}

\catcode`@=11
\newcount\@tempcntc
\def\@citex[#1]#2{\if@filesw\immediate\write\@auxout{\string\citation{#2}}\fi
  \@tempcnta\z@\@tempcntb\m@ne\def\@citea{}\@cite{\@for\@citeb:=#2\do
    {\@ifundefined
       {b@\@citeb}{\@citeo\@tempcntb\m@ne\@citea\def\@citea{,}{\bf ?}\@warning
       {Citation `\@citeb' on page \thepage \space undefined}}%
    {\setbox\z@\hbox{\global\@tempcntc0\csname b@\@citeb\endcsname\relax}%
     \ifnum\@tempcntc=\z@ \@citeo\@tempcntb\m@ne
       \@citea\def\@citea{,}\hbox{\csname b@\@citeb\endcsname}%
     \else
      \advance\@tempcntb\@ne
      \ifnum\@tempcntb=\@tempcntc
      \else\advance\@tempcntb\m@ne\@citeo
      \@tempcnta\@tempcntc\@tempcntb\@tempcntc\fi\fi}}\@citeo}{#1}}
\def\@citeo{\ifnum\@tempcnta>\@tempcntb\else\@citea\def\@citea{,}%
  \ifnum\@tempcnta=\@tempcntb\the\@tempcnta\else
   {\advance\@tempcnta\@ne\ifnum\@tempcnta=\@tempcntb \else \def\@citea{--}\fi
    \advance\@tempcnta\m@ne\the\@tempcnta\@citea\the\@tempcntb}\fi\fi}
\catcode`@=12
\newcommand{\la}{\lambda}
\newcommand{\La}{\Lambda}
\newcommand{\bP}{\bar{\Psi}_L}
\newcommand{\cD}{{\cal D}}
\newcommand{\cL}{{\cal L}}
\newcommand{\cO}{{\cal O}}
\newcommand{\cM}{{\cal M}}
\relax
\begin{document}
\ls{1}
\par \vskip .05in
\FERMILABPub{95/322--T}
\begin{titlepage}
\begin{flushright}
FERMILAB--PUB--95/322--T\\
TUM-HEP-228/95\\
October, 1995\\
Submitted to {\em Phys. Rev. }{\bf D}
\end{flushright}
\vskip 0.5 in
\begin{center}
{\large \bf GIM Violation and New\\
Dynamics of the Third Generation}
 \end{center}
  \par \vskip .1in \noindent
\begin{center}
{\bf Gerhard Buchalla$^a$\footnote{ Electronic address:
buchalla@fnth20.fnal.gov}, Gustavo Burdman$^a$\footnote{ Electronic address:
burdman@fnth22.fnal.gov}
\\ C. T. Hill$^a$\footnote{ Electronic address:
hill@fnal.gov}, Dimitris Kominis$^b$\footnote{ Electronic address:
kominis@physik.tu-muenchen.de}
  \par \vskip .2in \noindent
{$^a$\em Fermi National Accelerator Laboratory
P.O. Box 500, \\ Batavia, Illinois, 60510, USA. }\\
\vskip 0.1in
{$^b$\em Institut f\"ur Theoretische Physik, Technische Universit\"at
M\"unchen, \\ James-Franck-Strasse, 85748 Garching, Germany.}}
\end{center}
  \par \vskip .2in \noindent
\begin{center}{\large Abstract}\end{center}
\par \vskip .01in
\begin{quote}
In strong dynamical schemes for electroweak symmetry breaking
the third generation must be treated in a special manner,
owing to the heavy top quark.  This potentially  leads
to new flavor physics involving the members of the third
generation in concert with the adjoining generations, with
potential novel effects in beauty and charm physics.
We give a general discussion and formulation of this kind
of physics, abstracted largely from Topcolor models which
we elaborate in detail.  We identify sensitive channels for
such new physics
accessible to current and future experiments.
 \end{quote}
 \par \vskip .02in \noindent

\vfill
\end{titlepage}
\def\baselinestretch{1.6}
\tiny
\normalsize

\ls{1} %%%%%%%% Modifies Spacing to 1 %%%%%%%%5

\vskip .1in
\noindent
{\Large\bf I. Introduction}
\vskip .1in

\noindent
The problem of understanding the origin
of electroweak symmetry breaking is far from solved.  The fermion
Dirac masses arise in conjunction with the electroweak symmetry
breaking
since the left--handed members have weak isospin $I=1/2$ while
the right--handed members have $I=0$.  While the lightest quarks and
leptons can be regarded as perturbative spectators to the electroweak
dynamics, the very massive top quark suggests that it, and thus the
third generation, are potentially enjoying a more intimate role
in the electroweak dynamics and/or horizontal symmetry breaking.
A potential implication of this is the possibility that
there exist new fermion
interactions that do not treat the generations
in an egalitarian manner {\em at the electroweak scale}.  If there
are
dynamical distinctions between the generations
at the electroweak scale, then there is the possibility of new
observable phenomena which violate the GIM structure of the Standard
Model interactions. An example is
any description in which electroweak symmetry breaking is dynamical,
in analogy
with chiral symmetry breaking in QCD, such
as Technicolor (TC) and extended Technicolor (ETC) \cite{TC}.
These approaches require
special treatment of the large top quark mass generation.
Various mechanisms for a large
top mass have been proposed, including walking TC \cite{WETC},
subcritical amplification \cite{Ap}, two--scale
Technicolor \cite{EL}, and Topcolor \cite{TopC0,TopC1,TopC2}.

In the present paper we will focus on Topcolor, because it is fairly
well defined within the context of the existing fermionic
generations, and has direct implications of the general kind
we wish to consider.  However, we view it as generic in the possible
new GIM violating effects that it generates. Thus, we use Topcolor
in the present paper as a generating mechanism for possible signals
of new physics that might arise in detailed observations of, mostly,
$b$ and $c$ quark weak processes.
Topcolor assumes that most of the top quark mass arises from
a $t\bar{t}$ condensate. Previously,
top quark condensation models tried to identify
all of the Electroweak Symmetry Breaking (ESB) with the formation of
a
dynamical top quark mass \cite{BHL}, but
this requires a very large scale for the new dynamics
$\Lambda\sim 10^{15}$ GeV and significant fine--tuning.
In Topcolor we assume naturalness, i.e.,
the scale of the new physics is $O(1)$ TeV, and thus we estimate the
decay constant of the associated top--pions
by using the Pagels--Stokar formula in Nambu--Jona--Lasinio
approximation
\cite{BHL}. This gives
\beq
f_\pi^2 = \frac{N_c}{16\pi^2} m_{c}^2
( \log\frac{\Lambda^2}{m_c^2} + k)
\eeq
where $m_c$ is the dynamical mass,
$k$ a constant of $O(1)$,
and $\Lambda$ the cut-off scale at which
the dynamical mass is rapidly going
to zero. This results in
$f_\pi\sim 50$ GeV, a decay constant too small to account for all of
the
electroweak
symmetry breaking, which requires $f_\pi= 174$ GeV.
Hence we must postulate that Topcolor is
occuring in tandem with some other mechanism that gives most of
the electroweak scale.  This means that the top--pions are not
the longitudinal $W$ and $Z$, but are separate, physically observable
objects.  The top--pions must thus be massive, so in addition to
Topcolor
the top quark must derive some of its mass (about $\sim 3\%$)
from the electroweak breaking, allowing $m_{\tilde{\pi}}\sim 200$
GeV.

``Topcolor assisted Technicolor,'' was sketched out in
ref.\cite{TopC1}.
The specific model presented in \cite{TopC1} was based upon
the gauge group $SU(3)_1\times SU(3)_2
\times U(1)_{Y1}\times U(1)_{Y2}
\times SU(2)_L $, where the strong
double $U(1)_{Yi}$ structure is required
to tilt the chiral condensate in the $t\bar{t}$ direction, and not
form
a $b\bar{b}$ condensate. We shall refer to schemes based upon this
gauge structure, containing an additional $U(1)$, as Topcolor~I
models. Potentially serious problems with the $T$
parameter can arise \cite{Chivukula} in this scheme owing
to the strongly coupled $U(1)$, but they are avoided
by judicious choice of representations in Technicolor, and
reasonably complete models have been constructed \cite{EL2}.

In the present paper we give a discussion of the dynamical
features of Topcolor~I models, building upon the
recent work of one of us \cite{Kominis}. One of our main goals is
to provide an effective Lagrangian for the full boundstate dynamics.
This provides a natural starting point for the discussion
of other potentially observable effects.  One intriguing result is
that
the $\theta$--term in Topcolor can be the origin of observed
CP--violation,
yielding the CKM phase in the standard model and a Jarlskog
determinant of the right magnitude.

The potentially observable effects we are interested in arise because
in the current basis of quarks
and leptons the third generation experiences new strong forces.
When we diagonalize the mass matrix to arrive at the mass basis
there will be induced flavor changing interactions.
Some of these have been previously discussed \cite{TopC1,Kominis}.
Effects like
$B\overline{B}$ mixing are potentially dangerous.
The first and second
generational mixing effects are suppressed because the third
generation is
somewhat isolated, and these effects involve high powers of small
mixing
angles.  Topcolor, to an extent, explains the suppression
of the $3\rightarrow 2,1$ mixing angles, though without further
assumptions
about the origin of generational structure it cannot distinguish
between
$1$ and $2$.

We will sketch how the Topcolor scheme can impose textures
upon the mass matrix which has important consequences for
observable processes.  Textures are inevitable when there are  gauge
quantum numbers that distinguish generations.  A chiral--triangular
texture seems to emerge as a natural possibility, and this
can suppress dangerous processes such as $B\overline{B}$ mixing.

The Topcolor I models will be discussed in the context of the
implications
for GIM violation and new flavor physics. Here the additional
$U(1)$
gives rise to semileptonic processes of interest.  The model in
this truncated sector is somewhat akin to Holdom's generational $Z'$
model, with similar implications \cite{Holdom}.
We will also present a
class of models, Topcolor~II (essentially based upon \cite{TopC0}),
built upon the gauge group
$SU(3)_Q\times SU(3)_1\times SU(3)_2
\times U(1)_{Y} \times SU(2)_L $, where there is only the
conventional
$U(1)_{Y}$, and no strong additional $U(1)$.  These models have
several desirable features and have a rather intriguing anomaly
cancellation solution in which the $(c, s)_{L,R}$ doublets are
treated differently under the strong  $SU(3)_1\times SU(3)_2$
structure.
This leads to potentially interesting implications for charm physics
in
sensitive experiments.

Section IV. of the paper deals with the phenomenological
signatures of the new dynamics. It can be read  independently
of the theoretical discussions.  We identify interesting sensitivities
in some nonleptonic process such as $B\overline{B}$ and
$D\overline{D}$ mixing, and radiative processes such as $b\rightarrow s\gamma$.
However, we find that, in general, the strong dynamics at
the TeV scale is difficult to  observe in nonleptonic modes.
On the other hand, the semileptonic modes we identify are
interesting and sensitive to the $Z'$ of the Topcolor I schemes
(as well as in other $Z'$ schemes).  In general, Topcolor dynamics remains
viable at the current level of sensitivity and poses interesting experimental
challenges in high statistics heavy flavor experiments.

\vskip .3in
\noindent
{\Large\bf II. Topcolor Dynamics }
\vskip .1in
\noindent
\noindent {\large\em A.~Models with a Strong $U(1)$ to Tilt the
Condensate
(Topcolor~I).}
\vskip .1in
\noindent

We consider the possibility that the
top quark mass is large because
it is a combination
of a {\em dynamical condensate
component}, $(1-\epsilon)m_t$,
generated by a new strong dynamics,
together with a small {\em fundamental component},
$\epsilon m_t$, i.e, $\epsilon<<1$, generated by an
extended Technicolor (ETC) or Higgs sector.
The new strong dynamics is assumed to
be chiral--critically strong but
spontaneously broken, perhaps by TC itself, at the
scale $\sim 1$ TeV, and it is
coupled preferentially to the third
generation.
The new strong dynamics therefore
occurs primarily in interactions
that involve $\overline{t}t\overline{t}t$,
$\overline{t}t\overline{b}b$, and
$\overline{b}b\overline{b}b$,
while the ETC
interactions of the form $\overline{t}t\overline{Q}Q$, where
$Q$ is a Techniquark,
are relatively
feeble.

Our basic assumptions
leave little freedom of choice in
the new dynamics. We assume a new
class of Technicolor models
incorporating ``Topcolor'' (TopC).
In TopC~I the dynamics at the $\sim 1$ TeV scale
involves the following structure
(or a generalization thereof):
\beq
SU(3)_1\times SU(3)_2
\times U(1)_{Y1}\times U(1)_{Y2}
\times SU(2)_L \rightarrow
SU(3)_{QCD}\times U(1)_{EM}
\eeq
where
$SU(3)_1\times U(1)_{Y1}$
($SU(3)_2\times U(1)_{Y2}$)
generally couples preferentially
to the third (first and
second) generations.  The
$U(1)_{Yi}$
are just strongly rescaled
versions of
electroweak  $U(1)_{Y}$.
We remark that employing a new $SU(2)_{L,R}$
strong interaction in the third generation is also thinkable, but may
be problematic due to potentially large instanton effects that
violate $B+L$.  We will not explore this latter possibility
further.

The fermions are then assigned
$(SU(3)_1, SU(3)_2, {Y_1}, {Y_2}$) quantum numbers in the following
way:
\bea
(t,b)_L \;\;   &\sim  & (3,1,{1}/{3},0) \qquad \qquad
(t,b)_R \sim \left(3,1,({4}/{3},-{2}/{3}),0\right) \\ \nonumber
(\nu_\tau,\tau)_L &\sim & (1,1,-1,0) \qquad \qquad
\tau_R \sim \left(1,1,-2,0\right) \\ \nonumber
  & & \\ \nonumber
(u,d)_L,\;\;  (c,s)_L & \sim & (1,3,0,{1}/{3}) \qquad \qquad
(u,d)_R, \;\; (c,s)_R \sim \left(1,3,0,({4}/{3},-{2}/{3})\right) \\
\nonumber
(\nu, \ell)_L\;\; \ell = e,\mu & \sim & (1,1,0,-1) \qquad \qquad
\ell_R \sim \left(1,1,0,-2\right)
\eea
Topcolor must be broken, which we will assume is
accomplished through an (effective) scalar field:
\beq
\Phi \sim  (3,\bar{3}, y, -y) \label{phi_q}
\eeq
When $\Phi$ develops a Vacuum Expectation Value (VEV),
it produces the  simultaneous symmetry breaking
\beq
SU(3)_1\times SU(3)_2 \rightarrow
SU(3)_{QCD}\qquad
\makebox{ and}
\qquad
U(1)_{Y1}\times U(1)_{Y2}
\rightarrow  U(1)_{Y}
\label{sym_bre}
\eeq
The choice of $y$ will be specified below.

$SU(3)_1\times U(1)_{Y1}$ is assumed to be
strong enough to form
chiral condensates which will
naturally be tilted in the top
quark direction by the $U(1)_{Y1}$ couplings.
The theory is assumed to spontaneously break down to ordinary QCD
$\times U(1)_{Y}$ at a scale of $\sim 1$~TeV, before it becomes confining.
The isospin splitting that permits the formation of a $\VEV{\overline{t}t}$
condensate but disallows the $\VEV{\overline{b}b}$ condensate is due to the
$U(1)_{Yi}$ couplings. Since they are both larger than the ordinary hypercharge
gauge coupling, no significant fine--tuning is needed
in principle to achieve this symmetry
breaking pattern.
The $b$--quark mass in this scheme
is then an interesting issue,
arising from a combination
of ETC effects and instantons
in  $SU(3)_1$. The $\theta$--term
in $SU(3)_1$ may manifest itself as
the CP--violating phase in the CKM matrix.
Above all, the new spectroscopy
of such a system
should begin to materialize
indirectly
in the third generation
(e.g., in $Z\rightarrow \overline{b}b$)
or perhaps at the Tevatron in top
and bottom quark production.
The symmetry breaking pattern outlined above will generically give rise to
three (pseudo)--Nambu--Goldstone bosons $\tilde{\pi}^a$,
or``top-pions'', near the top mass scale. If the topcolor scale is of the order
of 1~TeV, the top-pions will
have a decay constant
of $f_\pi  \approx 50$ GeV, and a
strong coupling
given by a Goldberger--Treiman
relation,
$g_{tb\pi} \approx m_t/\sqrt{2}f_\pi\approx 2.5$,
potentially
observable in
$\tilde{\pi}^+\rightarrow t + \overline{b}$
if $m_{\tilde{\pi}} > m_t + m_b$.

We assume that ESB can be primarily driven
by a Higgs sector or Technicolor, with gauge group $G_{TC}$.
Technicolor can also provide
condensates which generate the
breaking of Topcolor
to QCD and $U(1)_Y$, although this can also be done by a Higgs field.
The coupling constants (gauge
fields) of
$SU(3)_1\times SU(3)_2$  are
respectively
$h_1$ and $h_2$ ($A^A_{1\mu}$
and $A^A_{2\mu}$)
while for $U(1)_{Y1}\times U(1)_{Y2}$
they
are respectively  ${q}_1$ and $q_2$,
$(B_{1\mu}, B_{2\mu})$.
The $U(1)_{Yi}$ fermion couplings are
then $q_i\frac{Yi}{2}$, where $Y1, Y2$
are the charges of the fermions under $U(1)_{Y1}, U(1)_{Y2}$ respectively.
A $(3,\overline{3})\times
(y,-y)$
Techni--condensate (or Higgs field) breaks $SU(3)_1\times SU(3)_2
\times U(1)_{Y1}\times U(1)_{Y2}
\rightarrow SU(3)_{QCD}\times  U(1)_Y$
at a scale
$\Lambda \gta 240$ GeV, or it
fully breaks $SU(3)_1\times SU(3)_2
\times U(1)_{Y1}\times U(1)_{Y2}\times SU(2)_L
\rightarrow SU(3)_{QCD}\times  U(1)_{EM}$
at the scale $\Lambda_{TC}= 240$ GeV.
Either scenario typically
leaves a {\em residual global symmetry},
$SU(3)'\times U(1)'$,
implying a degenerate, massive
color octet of ``colorons,'' $B_\mu^A$,
and a singlet heavy
$Z'_{\mu}$.  The gluon $A_\mu^A$
and coloron $B_\mu^A$ (the SM $U(1)_Y$
field
$B_\mu$ and the $U(1)'$ field
$Z'_\mu$), are then defined by
orthogonal
rotations with mixing angle
$\theta$ ($\theta'$):
\bea
& &
h_1\sin\theta = g_3;\qquad
h_2\cos\theta = g_3;\qquad
\cot\theta = h_1/h_2;\qquad
\frac{1}{g_3^2} = \frac{1}{h_1^2} +
 \frac{1}{h_2^2} ;
\nonumber \\
& &
q_1\sin\theta' = g_1;\qquad
q_2\cos\theta' = g_1;\qquad
\cot\theta' = q_1/q_2;\qquad \nonumber \\
& &
\qquad\qquad\qquad\qquad\qquad\frac{1}{g_1^2} = \frac{1}{q_1^2} +
 \frac{1}{q_2^2} ;
\eea
and $g_3$ ($g_1$) is the QCD ($U(1)_Y$)
coupling constant at $\Lambda_{TC}$.
%The overall scale factor $y$ can be adjusted to unity
%by appropriate rescaling of the $q_i$, and
%arises because the overall scale of the abelian $U(1)$ couplings is
%not fixed.
We ultimately demand $\cot\theta \gg 1$
and  $\cot\theta' \gg 1$
to select the top quark direction for
 condensation.
The masses of the degenerate octet of
colorons and $Z'$ are given
by $M_B\approx g_3\Lambda/\sin\theta\cos\theta$,
$ M_{Z'} \approx y g_1\Lambda/\sin\theta'\cos\theta'$.
The usual QCD gluonic ($U(1)_Y$ electroweak)
interactions
are obtained for any quarks that carry
either $SU(3)_1$
or $SU(3)_2$ triplet quantum numbers
(or $U(1)_{Yi}$ charges).

The coupling of the new heavy bosons $Z'$ and $B^A$ to fermions
is then given by
\begin{equation}\label{lzb}
{\cal L}_{Z'}=g_1\cot\theta' Z'\cdot J_{Z'} \qquad
{\cal L}_{B}=g_3\cot\theta B^A\cdot J^A_{B}
\end{equation}
where the currents $J_{Z'}$ and $J_B$ in general involve all three
generations of fermions
\begin{equation}\label{jzb123}
J_{Z'}=J_{Z',1}+J_{Z',2}+J_{Z',3}\qquad
J_B=J_{B,1}+J_{B,2}+J_{B,3}
\end{equation}
For the third generation the currents read explicitly
(in a weak eigenbasis):
\begin{eqnarray}\label{jz3}
J^\mu_{Z',3} &=& \frac{1}{6}\bar t_L\gamma^\mu t_L+
\frac{1}{6}\bar b_L\gamma^\mu b_L+\frac{2}{3}\bar t_R\gamma^\mu t_R
-\frac{1}{3}\bar b_R\gamma^\mu b_R \\ \nonumber
& & -\frac{1}{2}\bar\nu_{\tau L}\gamma^\mu \nu_{\tau L}
-\frac{1}{2}\bar\tau_L\gamma^\mu\tau_L-\bar\tau_R\gamma^\mu\tau_R
\end{eqnarray}
\begin{equation}\label{jb3}
J^{A,\mu}_{B,3}=\bar t\gamma^\mu\frac{\lambda^A}{2}t+
\bar b\gamma^\mu\frac{\lambda^A}{2}b
\end{equation}
where $\lambda^A$ is a Gell-Mann matrix acting on color indices.
For the first two generations the expressions are similar, except for
a suppression factor of $-\tan^2\theta'$ ($-\tan^2\theta$)
\begin{equation}\label{jz2}
J^\mu_{Z',2}=-\tan^2\theta'\left(\frac{1}{6}\bar c_L\gamma^\mu c_L+
\frac{1}{6}\bar s_L\gamma^\mu s_L +\ldots \right)
\end{equation}
\begin{equation}\label{jb2}
J^\mu_{B,2}=-\tan^2\theta\left(\bar
c\gamma^\mu\frac{\lambda^A}{2}c+
\bar s\gamma^\mu\frac{\lambda^A}{2}s \right)
\end{equation}
with corresponding formulae applying to the first generation.
Integrating out the heavy bosons $Z'$ and $B$, these couplings give
rise
to effective low energy four fermion interactions, which can in
general be written as
\begin{equation}\label{lzbeff}
{\cal L}_{eff,Z'}=-\frac{2\pi\kappa_1}{M^2_{Z'}} J_{Z'}\cdot J_{Z'}
\qquad
{\cal L}_{eff,B}=-\frac{2\pi\kappa}{M^2_B} J^A_B\cdot J^A_B
\end{equation}
where
\begin{equation}\label{kk1def}
\kappa_1=\frac{g^2_1\cot^2\theta'}{4\pi} \qquad
\kappa=\frac{g^2_3\cot^2\theta}{4\pi}
\end{equation}

The effective Topcolor interaction
of the third generation takes the form:
\bea
{\cal{L}}'_{TopC} & =  & -\frac{2\pi\kappa}{M_B^2}\left(
\bar{t}\gamma_\mu \frac{\lambda^A}{2} t +
 \bar{b}\gamma_\mu \frac{\lambda^A}{2} b
\right)
\left(
\bar{t}\gamma^\mu \frac{\lambda^A}{2} t +
 \bar{b}\gamma^\mu \frac{\lambda^A}{2} b
\right) .\label{topc_in}
\eea
This interaction is attractive in the color-singlet $\bar{t}t$ and $\bar{b}b$
channels and
invariant under color $SU(3)$
and $SU(2)_L\times SU(2)_R
\times U(1) \times U(1)$ where $SU(2)_R$ is the custodial
symmetry of the electroweak interactions.

In addition to the Topcolor interaction, we have the $U(1)_{Y1}$
interaction (which breaks custodial $SU(2)_R$):
\bz
{\cal{L}}'_{Y1} =   -\frac{2\pi\kappa_1}{M^2_{Z'}}\left(
\frac{1}{6}\bar{\psi}_L\gamma_\mu \psi_L +
\frac{2}{3}\bar{t}_R\gamma_\mu t_R
-\frac{1}{3}\bar{b}_R\gamma_\mu  b_R -
\frac{1}{2}\bar{\ell}_L\gamma_\mu \ell_L -
\bar{\tau}_R\gamma_\mu \tau_R \right)^2
\label{u1_in}
\ez
where $\psi_L = (t,b)_L$, $\ell_L = (\nu_\tau,\tau)_L$
and $\kappa_1$ is assumed to be $O(1)$. (A small value for $\kappa_1$ would
signify fine-tuning and may be phenomenologically undesirable.)

The attractive TopC interaction, for sufficiently large
$\kappa$, can trigger the formation of a
low energy
condensate, $\VEV{\overline{t}t + \overline{b}b}$, which
would break $SU(2)_L\times SU(2)_R\times  U(1)_Y
\rightarrow U(1)\times SU(2)_{c}$, where $SU(2)_{c}$ is
a global custodial symmetry. On the
other hand,  the $U(1)_{Y1}$ force is attractive in the
${\overline{t}t}$
channel and repulsive in the ${\overline{b}b}$ channel.  Thus, we can
have in concert critical and subcritical values of the combinations:
\beq
\kappa + \frac{2\,\kappa_1}{9N_c} > \kappa_{crit} ;
\qquad
\kappa_{crit} > \kappa - \frac{\kappa_1}{9N_c}.
\label{crit_con}
\eeq
Here $N_c$ is the number of colors. It should be mentioned that this analysis
is performed in the context of a large-$N_c$ approximation. The leading
isospin-breaking effects are kept even though they are ${\cal O}(1/N_c)$. The
critical coupling, in this approximation, is given by $\kappa_{crit} =
2\pi/N_c$. In what follows, we will not make explicit the $N_c$ dependence, but
rather take $N_c=3$.
We would expect the cut--off for integrals
in the usual Nambu--Jona-Lasinio (NJL) gap equation for $SU(3)_{TopC}$
($U(1)_{Y1}$) to be $\sim M_B$ ($\sim M_{Z'}$). Hence, these
relations define
criticality conditions irrespective of $M_{Z'}/M_B$.
This leads to ``tilted'' gap equations in which the top
quark acquires a constituent mass, while the $b$ quark
remains massless.  Given that both $\kappa$ and $ \kappa_1$
are large there is no particular fine--tuning occuring here,
only ``rough--tuning'' of the desired tilted configuration.
Of course, the NJL approximation is crude, but as long as the
associated
phase transitions of the real strongly
coupled theory are approximately second order, analogous
rough--tuning in the full theory is possible.

The full phase diagram of the model is shown in Fig.~1.
The criticality conditions (\ref{crit_con}) define the allowed region
in
the $\k_1$--$\k$ plane in the form of the two straight solid lines
intersecting
at $(\k_1=0,\k=\k_{crit})$. To the left of these lines lies the
symmetric
phase, in between them the region where only a $\VEV{\bar{t}t}$
condensate forms
and to the right of them the phase where both $\VEV{\bar{t}t}$ and
$\VEV{\bar{b}b}$ condensates arise. The horizontal line marks the
region above
which  $\k_1$ makes the $U(1)_{Y1}$ interaction strong enough to
produce
a $\VEV{\bar{\t}\t}$ condensate. (This line is meant
only as an indication, as
the fermion-bubble (large-$N_c$) approximation, which we use, evidently fails
for leptons.)
There is an additional constraint from
the measurement of $\G(Z\to\t^+\t^-)$, confining the allowed region
to the one below the solid curve. This curve corresponds to a $2\sigma$
discrepancy between the
topcolor prediction (computed to lowest non-trivial order in the coupling
$\kappa_1$) and the measured value of this width.
Note that the known value of the
top quark mass determines the cutoff $M_B$ in terms of
$\kappa$ and $\kappa_1$. This is illustrated by the slanted lines which
represent curves of constant $M_B$. In this figure,
the $Z^{\prime}$ boson mass is taken to be equal to $M_B$.
In the allowed
region a top condensate alone forms. The
constraints favor
a strong $SU(3)_{TopC}$ coupling and a relatively weaker $U(1)_{Y1}$
coupling.

We note that recently Appelquist and Evans have proposed a scheme in
which
the tilting interaction is a nonabelian gauge group, rather than a
$U(1)$
\cite{Appel_evans}.
This has the advantage of asymptotic freedom in the tilting
interaction
immediately above the scale of TopC condensation.

\vskip 0.3in
\noindent
{\large\em B.~Anomaly--Free Model Without a Strong $U(1)$
(Topcolor~II).}
\vskip .1in
\noindent

The strong $U(1)$ is present in the previous scheme to avoid
a degenerate $\VEV{\bar{t}t}$ with $\VEV{\bar{b}b}$.  However,
we can give a model in which there is: (i) a Topcolor $SU(3)$
group but (ii) no strong $U(1)$ with (iii) an anomaly-free
representation content. In fact the original
model of \cite{TopC0} was of this form, introducing a
new quark of charge $-1/3$.  Let us consider a
generalization of this scheme which consists of the gauge structure
$SU(3)_Q\times SU(3)_1\times SU(3)_2 \times U(1)_{Y}\times SU(2)_{L}$.
%at a scale  $ \Lambda $.
We require an additional triplet of fermion
fields $(Q_R^a)$ transforming as $(3,3,1)$
and $Q_L^{\dot{a}}$ transforming as $(3,1,3)$ under
the $SU(3)_Q\times SU(3)_1\times SU(3)_2$.

The fermions are then assigned the following quantum numbers
in $SU(2)\times SU(3)_Q \times SU(3)_1\times SU(3)_2\times U(1)_Y $:
\bea
(t,b)_L \;\;  (c,s)_L &\sim & (2,1,3,1) \qquad Y=1/3 \\ \nonumber
(t)_R &\sim & (1,1,3,1) \qquad Y=4/3;\qquad \\\nonumber
(Q)_R &\sim  & (1,3,3,1) \qquad Y=0\\ \nonumber
  & & \\ \nonumber
(u,d)_L &\sim & (2,1,1,3) \qquad Y=1/3 \\ \nonumber
(u,d)_R \;\; (c,s)_R &\sim & (1,1,1,3) \qquad Y=(4/3,-2/3) \\
\nonumber
(\nu, \ell)_L\;\; \ell = e,\mu,\tau &\sim & (2,1,1,1) \qquad Y=-1;
\qquad
\\ \nonumber (\ell)_R &\sim &  (1,1,1,1) \qquad Y=-2 \\ \nonumber
b_R & \sim & (1,1,1,3) \qquad Y= 2/3;   \qquad
\\ \nonumber  (Q)_L
& \sim & (1,3,1,3) \qquad Y= 0;
\eea
 Thus, the $Q$ fields are electrically
neutral.  One can verify that this assignment is anomaly free.

The $SU(3)_Q$
confines and
forms a $\VEV{\bar{Q}Q}$ condensate which acts
like the $\Phi$ field and breaks the Topcolor
group down to QCD dynamically.    We assume that $Q$ is then
decoupled from the
low energy spectrum by its large constituent mass.  There is only a
lone
$U(1)$ Nambu--Goldstone boson  $\sim {\bar{Q}\gamma^5 Q}$
which acquires a large mass by $SU(3)_Q$
instantons.

The $SU(3)_1 $ is chiral-critical, and a condensate forms
which defines the $\VEV{\bar{t}t}$ direction spontaneously.
If we
turn off the $SU(2)\times U(1)_Y$ and Higgs--Yukawa
couplings, then the strongly coupled
$SU(3)_1$ sector has an $SU(4)_L\times U(1)_R\times U(1)_L$ global
chiral symmetry.  Let us define $\Psi_L = (t,b,c,s)_L$. The
effect of $SU(3)_1$ after integrating out the massive colorons and
$Q$ fields is a strong 4-fermion, NJL--like interaction of the form
\bea
G(\overline{\Psi}_L^i t_R)(\overline{t}_R \Psi_{Li})
& \rightarrow &(\overline{\Psi}_{L}^i t_R F_i + h.c.) - G^{-1}F^\dagger F
\nonumber
\\
& = &  \left[(\overline{T}_L^i t_R)H_i  + (\overline{C}_L^i t_R)K_i
+ h.c. \right] -
G^{-1} (H^\dagger H + K^\dagger K)
\eea
where we indicate the factorization into a composite field
4-plet under $SU(4)_L$, $F^i$.  We further decompose $F$ into
doublets
$F=(H,K)$.
By definition, $H$ acquires a VEV giving the top mass, while the
remaining components of $H$ are a massive neutral Higgs-like
$\sigma$ boson,  and a triplet of
top-pions as before. Here the novelty
is that $K$ will be, at this stage, a completely massless set of
NGB's,
``charm-top--pions.''
When $SU(2)_L\times U(1)$ and the Yukawa interactions to the
effective Higgs field are
switched on, the top-pions and charm-top--pions all become massive.
These will then mediate strong interactions, but which are distinctly
nonleptonic in the present scheme. We discuss its phenomenological
consequences, mainly for $D^0-\bar{D}^0$ mixing, in section~IV.(B).

\vskip .1in
\noindent
{\large\em C.~Triangular Textures }
\vskip .1in
\noindent
The texture of the fermion mass matrices will generally
be controlled by the symmetry breaking pattern of a horizontal
symmetry.  In the present case we are specifying a residual
Topcolor symmetry, presumably  subsequent to some
initial breaking at some
scale $\Lambda$, large compared to Topcolor, e.g., the third
generation fermions in Model~I have different Topcolor assignments than
do the
second and first generation fermions. Thus the texture will depend
in some way upon the breaking of Topcolor.

Let us study a fundamental Higgs boson, which ultimately
breaks $SU(2)_L\times U(1)_Y$,
together with an effective field $\Phi$  breaking Topcolor
as in eq.(\ref{sym_bre}).  We must now specify the full Topcolor
charges
of these fields. As an example, under
$SU(3)_1\times SU(3)_2 \times U(1)_{Y1}\times U(1)_{Y2}\times
SU(2)_L$
let us choose:
\beq
\Phi \sim (3,\bar{3}, \frac{1}{3}, -\frac{1}{3}, 0)
\qquad
H \sim (1,1,0, -1, \half)
\eeq
The effective couplings to fermions that generate mass terms in the
up sector are of the form
\by
-{\cal L}_{{\cal M}_U}&=&m_0 \bar{t}_Lt_R +c_{33}\bar{T}_Lt_R
H\fr{det\F^\dagger}{\L^3}+
c_{32}\bar{T}_L c_R H\fr{\F}{\L} + c_{31}\bar{T}_L u_R
H\fr{\F}{\L}\nn\\
& &+c_{23}\bar{C}_L t_R H \F^\dagger
\fr{det\F^\dagger}{\L^4}+c_{22}\bar{C}_L c_R
H + c_{21}\bar{C}_L u_R H  \label{lag_mass}\\
& & + c_{13}\bar{F}_L t_R H \F^\dagger
\fr{det\F^\dagger}{\L^4}+c_{12}\bar{F}_L c_RH
+c_{11}\bar{F}_L u_R H  + {\rm h.c.} \nn
\ey
Here $T=(t,b)$, $C=(c,s)$ and $F=(u,d)$.  The mass $m_0$ is
the dynamical condensate top mass.
Furthermore, $det\Phi$ is defined by
\bz
det \Phi \equiv \frac{1}{6}\e_{ijk}\e_{lmn}\Phi_{il}\Phi_{jm}\Phi_{kn}
\label{phi_det}
\ez
where in $\Phi_{rs}$ the first (second) index
refers to $SU(3)_1$ ($SU(3)_2$).
The matrix elements now require factors
of $\Phi$ to connect the third with the
first or second generation color indices. The down quark
and lepton mass matrices are generated by couplings analogous to
(\ref{lag_mass}).

To see what kinds of textures can arise naturally,
let us assume that the ratio $\Phi/\Lambda$ is small, O($\epsilon$).
The field $H$ acquires a VEV of $v$.
Then the resulting mass  matrix is approximately triangular:
\bea
&& \left( \begin{array}{ccc}
c_{11}v &  c_{12}v & \sim 0    \cr
 c_{21}v & c_{22}v  &\sim 0    \cr
c_{31}O(\epsilon)v & c_{32}O(\epsilon)v & \sim m_0 + O(\epsilon^3)v
\cr
 \end{array}\right)\label{m_trian}
\eea
where we have kept only terms of $\cal O (\e)$ or larger.

This is a triangular matrix (up to the $c_{12}$ term).
When it is written in the form
$U_L {\cal D} U^{\dagger}_R$ with $U_L$ and $U_R$ unitary and ${\cal
D}$
positive diagonal,
there automatically result restrictions on $U_L$ and $U_R$.  In the
present case,
the elements $U^{3,i}_L$ and $U^{i,3}_L$ are vanishing for
$i\neq 3$ , while the elements of $U_R$ are not constrained
by triangularity.
Analogously, in the down quark sector $D^{i,3}_L=D^{3,i}_L=0$ for
$i\neq 3$
with $D_{R}$ unrestricted.
The situation is reversed when the opposite corner elements are
small,
which can be achieved by choosing $H \sim (1,1, -1, 0, \half)$.

These restrictions on the quark rotation matrices have important
phenomenological consequences.
For instance, in the process $B^0\rightarrow \overline{B^0}$ there are
potentially
large contributions from top-pion and coloron exchange.  However, as
we show in
Section~IV.(B), these contributions are proportional to the
product $D^{3,1}_L D^{3,1}_R$.  The same occurs in $D^0-\bar{D}^0$
mixing, where
the effect goes as products involving $U_L$ and $U_R$ off-diagonal
elements.
Therefore, triangularity can naturally select
these products to be small.

Selection rules will be a general consequence in models where the
generations have different  gauge quantum numbers above some scale.
The precise selection rules depend upon the particular symmetry
breaking
that occurs. This example is merely illustrative of the
systematic effects that can occur in such schemes.
The model of Ref.~\cite{EL2} provides a specific realization
of triangular textures in the context of ETC, as pointed out
in \cite{lane_pc}.

\vskip .3in
\noindent
{\Large\bf III.  Effective Lagrangian Analysis}
\vskip 0.1in
\noindent
{\large\em A. Low Energy Theory }
\vskip .1in
\noindent

Let us study a standard model Higgs boson
together with an effective Nambu--Jona-Lasinio  \cite{NJL}
mechanism arising from the
interactions of eqns.~(\ref{topc_in}) and  (\ref{u1_in}).
The standard model Higgs boson, $H$, is used
presently to simulate the effects of TC,
and its small Yukawa couplings
to the top and bottom quarks simulate the effects of ETC.  In
addition,
the four--fermion interaction is introduced to
simulate the effects of TopC (Model~I).

We can conveniently treat the dynamics of
this combined system  using the renormalization group
by writing a
Yukawa form of the four--fermion interactions, as defined at the
cut--off scale $\Lambda$, with  the help of a static auxiliary
Higgs field.

At the starting point we will have three boundstate
doublets: an ordinary Higgs $H$ whose
VEV drives electroweak symmetry breaking, a doublet $\phi_1$ whose
VEV
mainly gives a large mass to the top and another doublet $\phi_2$
coupling mainly to bottom \cite{Kominis}.
The effective Lagrangian at the cutoff
$\La$ (identified with $M_B \approx M_{Z'}$ of Section II.A) is
\begin{eqnarray}
\cL & = & \bar{\Psi}_L^{(3)} \phi_1 t_R + \bar{\Psi}_L^{(3)} \phi_2
b_R +
\kappa^{\prime} \La ^2 e^{i\theta} {\rm det}\, (\phi_1 \; \phi_2)+
{\rm  h.c.}  - \La_1^2
\phi_1^{\dagger} \phi_1 - \La_2^2 \phi_2^{\dagger} \phi_2 \nonumber
\\
 & & + D_{\mu}H^{\dagger} D^{\mu}H - M_H^2 H^{\dagger}H -
\frac{\la}{2}
(H^{\dagger}H)^2 + \left [\epsilon_U^{ij} \bP ^i H U_R^j
+ \epsilon_D^{ij} \bP ^i H^c D_R^j + {\rm h.c.} \right ] \nonumber \\
 & & + \cL_{gauge}
\label{e1}
\end{eqnarray}
where $\bP, U_R, D_R$ have the
obvious meaning, $\cL_{gauge}$ contains the gauge and fermion kinetic
terms,
and
$\La_{1,2}^2 = \La^2/\la_{1,2}^2$ satisfy  $\La_2 > \La_1$, with
\bz
\l_{1}^{2}=4\p\left(\k+\fr{2\;\k_1}{27}\right)\quad\quad ;
\quad\quad
\l_{2}^{2}=4\p\left(\k-\fr{\k_1}{27}\right).
\label{lam_defs}
\ez
The parameters $\kappa$ and $\kappa_1$ were defined in Section~II.A.
In (\ref{e1}), $i,j$ are generation indices.
The determinant term in the expression for the effective Lagrangian
%OR: in the same equation
arises from Topcolor instantons,
and $\theta$ is the strong Topcolor CP--phase.
This term is a bosonized form of a 't Hooft flavor determinant
$k e^{i\theta} {\rm det}(\bar{\Psi}_L^{(3)} \Psi_R^{(3)}) / \Lambda^2$, where
the constant $k$ is expected to be ${\cal O}(1)$.
These effects are similar to those in QCD that elevate the $\eta'$
mass;
we assume the QCD CP--phase is zeroed by, e.g., an exact Peccei-Quinn
symmetry.
Presently, the Topcolor $\theta$ angle will provide an origin for
CKM CP--violation.
The coefficient
$\kappa^{\prime}$ is related to $k$ through
\be
\kappa^{\prime} = \frac{k}{\lambda_1^2 \lambda_2^2}
\ee
and is thus small, as are
the matrices $\epsilon_{U,D}$. We shall write
\be
\epsilon_{U,D}^{ij} = \epsilon\, \eta_{U,D}^{ij}\, ,\;\;\;\;
\kappa^{\prime}
= -\epsilon \,\eta_k
\ee
and treat $\epsilon$ as small compared to unity.
The $\epsilon^{ij}_{U,D}$ will reflect the textures as considered in
section II.C.
The $\eta$'s are then
$\cO (1)$. Eq.(\ref{e1}) will be the starting point of our
investigations.

We now integrate out degrees of freedom with momenta between a scale
$\mu$ and the cutoff $\La$. This calculation is performed in the large-$N_c$
limit and cutting off the fermion loops at $\L$. The effective Lagrangian at
the scale $\mu$ has the form
\begin{eqnarray}
\cL & = & \cL_{gauge} + D_{\mu}H^{\dagger} D^{\mu}H - M_H^2
H^{\dagger}H
- \frac{\la}{2} (H^{\dagger}H)^2 + Z \,{\rm
tr}\,D_{\mu}\Sigma^{\dagger}
D^{\mu}\Sigma \nonumber \\
 & & - \La_1^2
\phi_1^{\dagger} \phi_1 - \La_2^2  \phi_2^{\dagger} \phi_2 +
\frac{3}{8\pi^2}(\La^2 - \mu^2) \,{\rm tr} \,\Sigma^{\dagger}\Sigma
\nonumber \\
 & & -\frac{3}{16\pi^2}\ln (\La^2/\mu^2) {\rm tr} (
 \Sigma^{\dagger}\Sigma)^2 + \left [\bP ^i \Sigma^{ij} \Psi_R^j +
{\rm h.c.} \right ] \nonumber \\
 & & + \left[\kappa^{\prime} \Lambda^2 e^{i \theta} {\rm det} \, (\phi_1 \;
 \phi_2)+ {\rm h.c.} \right ]
\label{e2}
\end{eqnarray}
where the $\Sigma^{ij}$ are matrices :
\be
\Sigma^{ij} = ( \phi_1 \delta ^{i3}\delta ^{j3} + \epsilon_U^{ij}H,
\quad\quad\phi_2 \delta ^{i3}\delta ^{j3} + \epsilon_D^{ij}H^c )
\label{sigma}
\ee
and the trace is taken over both generation and $SU(2)$ indices. The
constant $Z$ is given by
\be
Z=\frac{3}{16\pi^2} \ln (\La^2/\mu^2)
\ee
The mixing between $H$ and $\phi_1, \phi_2$ involves only
$\epsilon_{U,D}^{33}$ to first order in $\epsilon$. So, for the
scalar potential, we disregard the Higgs
couplings to the first two generations
and simplify $\Sigma$ into a $2\times 2$ matrix
\be
\Sigma = ( \phi_1 + \epsilon \eta_t H \quad\quad \phi_2 + \epsilon
\eta_b
H^c )
\ee
It is convenient to write $V$ in terms of $H$ and $\Sigma$ rather
than
$H$ and $\phi_i$. To $\cO (\epsilon)$,
we may also substitute det $\Sigma$ for det $(\phi_1, \;
\phi_2)$. Writing $F\equiv 3/(8\pi^2)$ and dropping
$\mu^2$ compared to $\La^2$, we have the following potential:
%\begin{eqnarray}
%V & = & M_H^2 H^{\dagger}H + \frac{\la}{2} (H^{\dagger}H)^2 + \La_1
%\phi_1^{\dagger} \phi_1 + \La_2 \phi_2^{\dagger} \phi_2 \nonumber \\
% & & + \epsilon \eta_k \La^2 (e^{i\theta} {\rm det}\, \Sigma + {\rm
%h.c.})
% - F \La^2 \, {\rm tr}\, \Sigma^{\dagger}\Sigma + Z \,{\rm tr}\, (
% \Sigma^{\dagger}\Sigma)^2
%\label{e3}
%\end{eqnarray}
\begin{eqnarray}
V & = & M_H^2 H^{\dagger}H + \frac{\la}{2} (H^{\dagger}H)^2 + \La_1^2
\Sigma_{i1}^*\Sigma_{i1} + \La_2^2 \Sigma_{i2}^* \Sigma_{i2}
- F \La^2 \,{\rm tr}\, \Sigma^{\dagger}\Sigma +Z \,{\rm tr}\,
(\Sigma^{\dagger}\Sigma)^2  \nonumber \\
 & & +\left [- \epsilon \,\La_1 ^2 \, \eta_t \Sigma_{i1}^* H_i +
 \epsilon \, \La_2 ^2 \, \eta_b \, \epsilon_{ij}\,
 \Sigma_{i2}^* H_j^*  + \epsilon \,\eta_k \,\La^2
 e^{i\theta} {\rm det}\, \Sigma + {\rm h.c.} \right ]
\label{e4}
\end{eqnarray}
where $\Sigma_{i1}$ is the first column of $\Sigma$ so that $\phi_{1i} =
\Sigma_{i1} - \epsilon \eta_t H_i$ etc. Note $H^c_i =
-\epsilon_{ij}H_j^*$. Without loss of generality we can take $\eta_t$
real and positive by appropriately choosing the phase of the field
$H$. The phase of $\eta_b$ can be absorbed in $\theta$ by an
appropriate
rotation of $b_R$.

To $\cO (\epsilon^0)$ the minimum of the potential is determined by
\begin{eqnarray}
\langle H^{\dagger}H \rangle  & \equiv & v_0^2  = -\frac{M_H^2}{\la}
\nonumber \\
\langle \Sigma_{i1}^*\Sigma_{i1} \rangle & \equiv & f_0^2 = \langle
\phi_1^{\dagger} \phi_1 \rangle \label{vevs} \\
\langle \Sigma_{i2}^*\Sigma_{i2} \rangle & = & 0
\nn
\end{eqnarray}
where $f_0$ satisfies the equation
\be
\La_1^2 - F\La ^2 + F f_0^2 \ln (\La^2 / \mu ^2) = 0
\label{e5}
\ee
and the last condition in eq.(\ref{vevs}) results from the assumption
that $\La_2^2 - F \La^2 > 0$.
We can work out the minimization of the potential to $\cO
(\epsilon)$. We find
\be
\langle H \rangle = \left ( \begin{array}{c} v_0 + \epsilon v_1 \\ 0
\end{array} \right ) \mbox{\hspace{1.5cm}} \langle \Sigma \rangle =
\left(
\begin{array}{cc} f_0 + \epsilon f_1 & 0 \\ 0 & \epsilon f_2
\end{array} \right )
\label{e7}
\ee
where
\begin{eqnarray}
{\rm Re} \,v_1 & = & \frac{\La_1^2 \eta_t f_0}{2\la v_0^2} \\
{\rm Im} \,v_1 & = & \frac{v_0}{\La_1 ^2\eta_t f_0} {\rm Im}
\,(\eta_b \eta_k
\La^2 f_0 e^{i\theta} + \La_2^2 \eta_b f_2^*) \\
f_1 & = & \frac{\La_1^2 \eta_t v_0}{4Z f_0^2} \\
f_2 & = & \frac{\La_2^2 \eta_b v_0 - \La ^2 \eta_k e^{-i\theta}
  f_0}{\La_2^2 - F \La^2}
\label{e8}
\end{eqnarray}
Here we have performed an $SU(2)\times U(1)$ rotation to make $\Sigma_{11}$
real and $\Sigma_{21}$ equal to zero.
Note, in particular, that the vacuum aligns properly, provided
$\eta_t
\ne 0$. This statement is
in fact true to all orders in $\epsilon$.
Furthermore, if we ignore mixing with the first two generations,
the top and bottom quark masses are given by
\begin{eqnarray}
m_t & = & |f_0 + \epsilon f_1 | \\
m_b & = & \epsilon |f_2|
\label{masses}
\end{eqnarray}
Note that the denominator in eq.~(\ref{e8}) can be small, thus enhancing the
size of the $b$-quark mass \cite{Kominis}.

In order to obtain expressions for the scalar and pseudoscalar
masses we must diagonalize the potential in (\ref{e4}).
Let $\chi_i^{\prime}$ denote the two columns of $\Sigma$ and define
\be
\chi_1 = \sqrt{Z}
\chi_1^{\prime}\mbox{\hspace{1.5cm}};\mbox{\hspace{1.5cm}} \chi_2 =
\sqrt{Z} \chi_2^{\prime \,c}
\ee
Then the kinetic term for the scalars is simply
\be
\cL_{kin} = D_{\mu}H^{\dagger} D^{\mu}H + D_{\mu}\chi_1^{\dagger}
D^{\mu}\chi_1 + D_{\mu}\chi_2^{\dagger} D^{\mu}\chi_2
\ee
and the potential reads
\begin{eqnarray}
V & = & M_H^2 H^{\dagger}H + Z^{-1}(\La_1^2-F\La^2)\chi_1^{\dagger}
\chi_1 +  Z^{-1}(\La_2^2-F\La^2)\chi_2^{\dagger} \chi_2 \nonumber \\
 & & -\epsilon \, \La_1^2 \,Z^{-1/2} \eta_t (\chi_1^{\dagger} H +
H^{\dagger}
 \chi_1) + \epsilon \, \La_2^2 \, Z^{-1/2} \eta_b (\chi_2^{\dagger} H
+ H^{\dagger}
 \chi_2) \nonumber \\
 & & - \epsilon \La^2 \, Z^{-1} \eta_k (e^{-i\theta} \chi_1^{\dagger}
 \chi_2 + {\rm h.c.}) + \frac{\la}{2} ( H^{\dagger}H)^2 \nonumber \\
 & & + Z^{-1}\left[ (\chi_1^{\dagger}
\chi_1)^2 + (\chi_2^{\dagger} \chi_2)^2 + 2 (\chi_1^{\dagger}\chi_1)
(\chi_2^{\dagger} \chi_2)- 2 (\chi_1^{\dagger}\chi_2)
(\chi_2^{\dagger}
\chi_1) \right ]
\label{e11}
\end{eqnarray}
The VEVs of the fields $H, \chi_1$ and $\chi_2$ add in quadrature to give the
electroweak scale (so, for example, to lowest order $v_w=\sqrt{v_0^2+Zf_0^2} =
174$~GeV).
We diagonalize the potential to obtain the mass eigenstates. In the
charged sector we find, apart from the Goldstone bosons $w^{\pm}$
of electroweak
symmetry breaking, a pair of (complex-conjugate) pseudo-Goldstone
bosons
(the top-pions) of mass
\be
m_{\tilde{\pi}^{\pm}} ^2 = \epsilon \La_1^2 \eta_t (\frac{v_0}{Zf_0}
+
\frac{f_0}{v_0} )
\label{e16a}
\ee
(in agreement with expectations from e.g. current algebra)
and a pair of massive states
\be
m_{\tilde{H}^{\pm}}^2 = Z^{-1}(\La_2^2-F\La^2 + 2Zf_0^2) + \epsilon
\La_1^2
\eta_t \frac{v_0}{Zf_0}
\label{e16b}
\ee
The charged components of the doublets $H$ and $\chi_1$ mix with an
angle:
\be
\phi = \arctan (Z^{1/2}f_0/v_0)
\ee
to produce the mass eigenstates
$w^{\pm}$ and $\tilde{\pi}^{\pm}$ (recall $Z^{1/2}f_0$ is the
top-pion
``decay constant''). There are further mixings at $\cO (\epsilon)$
among all three doublets, which are also easily calculable. They will
determine some of the couplings of fermions to scalars which do not
occur at leading order (e.g. top-pion couplings to $b_R$ or
$\tilde{H}^{\pm}$ couplings to $t_R$).

In the neutral sector we find the following eigenvalues:
\be
\begin{array}{lcl}
m_z^2 = 0  & ; & {\rm Goldstone\;boson} \nonumber \\
m_{\tilde{\pi}^0}^2 = \epsilon \La_1^2 \eta_t (\frac{v_0}{Zf_0} +
\frac{f_0}{v_0} ) & ; & {\rm neutral\;top-pion}  \nonumber \\
m_{A}^2 = Z^{-1}(\La_2^2 - F\La^2) & ; & {\rm bound\;state} \;
\bar{b}\gamma^5 b \nonumber \\
m_{\tilde{H}^0}^2 = Z^{-1}(\La_2^2 - F\La^2) & ; & {\rm bound\;
state}\;
\bar{b} b \nonumber \\
m_{h_1}^2 = 2\la v_0^2 + \cO (\epsilon) & ; & {\rm standard \;Higgs}
\nonumber
\\
m_{h_2}^2 = 4 f_0^2 +  \cO (\epsilon) & ; & {\rm ``top-Higgs"}
\end{array}\label{scal_spect0}
\ee
To leading order, the only mixing occurs between the imaginary parts
of
the neutral components of $H$ and $\chi_1$ -- the mixing angle is
the same as in the case of the charged sector, as expected.
There are further
mixings at $\cO (\epsilon)$. Due to the $CP$-violating angle $\theta$
there are mixings between $CP$-even and $CP$-odd fields.

\vskip 0.1in
\noindent
{\large\em B. ~Fermion Mass Matrices and Mixings}
\vskip 0.1in
\noindent

We will now determine whether this low-energy structure gives rise to
realistic quark mixings and weak $CP$ violation. Consider first the
mass
matrix in the `top' sector:
\be
-\cM _U ^{ij} = f_t \delta ^{i3} \delta ^{j3} + \epsilon \eta_U ^{ij}
v_0
\ee
where $|f_t| \approx m_t$. This is diagonalized by unitary matrices
$U_L$
and $U_R$ such that
\be
U_L^{\dagger} \cM _U U_R = \cD \label{udiag}
\ee
where $\cD = {\rm diag} ( m_u, \;m_c, \; m_t )$. The matrix $U_L$
has then the following approximate form
\be
U_L = \left ( \begin{array}{ccc} \cos \phi & -\sin \phi & \epsilon
\beta
\\ \sin \phi & \cos \phi & \epsilon \beta ^{\prime} \\
-\epsilon (\beta ^*\cos \phi + \beta ^{\prime *} \sin \phi) &
\epsilon (\beta ^* \sin \phi - \beta ^{\prime *} \cos \phi) & 1
\end{array}
\right )
\label{ul}
\ee
where $\phi$ is of order 1 and
\begin{eqnarray}
\epsilon \beta & = & \epsilon \frac{v_0 \eta_U^{13}}{f_t} \\
\epsilon \beta ^{\prime} & = & \epsilon \frac{v_0 \eta_U^{23}}{f_t}
\end{eqnarray}
Since the matrix $\eta_U$ determines the masses of the lower
generation
up-type quarks,
we expect $\epsilon \,v_0 \,\eta_U ^{j3} \sim m_c$. Hence $\epsilon
\beta,
\epsilon \beta ^{\prime} \sim \cO (m_c / m_t)$.
Similarly, in the `bottom' sector, the mass matrix is
\be
-\cM _D ^{ij} = f_b \delta ^{i3} \delta ^{j3} + \epsilon \eta_D ^{ij}
v_0
\ee
where $f_b = \langle \phi_2 \rangle _2$. In principle $f_b$ is of order
$\epsilon$, but we will assume that the instanton or enhancement
effects are large enough to account for the large $b$-quark mass
relative to that of the other charge --1/3 quarks (see the remark following
eq.~(\ref{masses})).
That is, we assume that
$f_b \gg \epsilon v_0 \eta_D^{ij}$. We obtain, as above, the
corresponding unitary matrix $D_L$. Note the following: (i) we can
perform an arbitrary rotation in the space of the $d$ and $s$ quarks
to
make the corresponding angle $\phi$ (cf. eq.(\ref{ul})) equal to
zero; (ii) we would like to investigate whether the angle $\theta$
can
be the sole source of weak $CP$ violation. Hence we would like to
take
the matrices $\eta_U, \eta _D$ to be real. Then the matrix $U_L$
above
is real, but $D_L$ is not, because $f_b$ is complex (see
eq.(\ref{e8})). The matrix $D_L$ then has the form
%\be
%D_L = \left ( \begin{array}{ccc}  1 & -\epsilon \gamma & \alpha
%e^{i\delta}
%\\ \epsilon \gamma ^* & 1 & \alpha ^{\prime}  e^{i\delta} \\
%- \alpha  e^{-i\delta} & -\alpha ^{\prime}  e^{- i\delta} & 1 \end{array}
%\right )
%\label{dl}
\be
D_L = \left ( \begin{array}{ccc}  1 & 0 & \alpha
e^{i\delta}
\\ 0 & 1 & \alpha ^{\prime}  e^{i\delta} \\
- \alpha  e^{-i\delta} & -\alpha ^{\prime}  e^{- i\delta} & 1 \end{array}
\right )
\label{dl}
\ee
where $\alpha, \alpha^{\prime}$ are real (and small),
\be
\alpha = \epsilon \,\frac{v_0
  \eta_D^{13}}{|f_b|}\mbox{\hspace{1.5cm}};\mbox{\hspace{1.5cm}} \alpha
^{\prime} = \epsilon \,\frac{v_0  \eta_D^{23}}{|f_b|}
\ee
and the angle $\delta$ can be of order 1:
\be
\tan \delta = \tan {\rm Arg } f_b^* \approx \frac{-\La ^2 \eta_k f_0
\sin
  \theta}{\La_2^2 \eta_b v_0 - \La^2 \eta_k f_0 \cos \theta}
\ee
If the instanton effects are dominant, then $|\delta| \approx
|\theta|$. Note that, since the matrix $\eta_D$ determines the down-
and
strange-quark masses, we expect $\alpha, \alpha^{\prime}
\sim \cO (m_s/m_b)$, as
$|f_b| \approx m_b$. So $\alpha, \alpha^{\prime}$ are expected to be larger
than
the corresponding elements ($\epsilon \beta, \epsilon
\beta^{\prime}$) of $U_L$.

The Kobayashi-Maskawa matrix $V \equiv U_L ^{\dagger} D_L$ now reads
\be
V = \left ( \begin{array}{ccc} c_{\phi} + \cO (\epsilon) & s_{\phi} +
\cO (\epsilon) & c_{\phi} (\alpha e^{i\delta} - \epsilon \beta) + s_{\phi}
(\alpha ^{\prime} e^{i\delta} - \epsilon \beta ^{\prime}) \\
- s_{\phi} + \cO (\epsilon) & c_{\phi} + \cO (\epsilon) &  c_{\phi}
(\alpha ^{\prime} e^{i\delta} - \epsilon \beta ^{\prime}) - s_{\phi} (\alpha
e^{i\delta} - \epsilon \beta) \\
-\alpha e^{-i\delta} + \epsilon \beta & -\alpha ^{\prime} e^{-i\delta} +
\epsilon
\beta ^{\prime} & 1 \end{array} \right )
\label{ckm}
\ee
where $c_{\phi} \equiv \cos \phi, s_{\phi} \equiv \sin \phi$.
This successfully predicts $|V_{cb}| \approx |V_{ts}| \approx
m_s/m_b$, but does not  distinguish between the first two
generations, given that we do not incorporate any dynamics to do so.
The $\eta$ matrices will have to be such as to suppress $\alpha$
and give $\phi \approx \theta_c$, the Cabibbo angle. A certain
cancellation will still have to occur between the two terms in
$V_{ub}$
in order to bring it to the correct value. We also note that, with
$\epsilon \beta \sim m_c/m_t$, $\alpha^{\prime} \sim m_s/m_b$ and $\alpha$
small, one obtains
a Jarlskog parameter of the right size:
%for values of $\delta$ of order one,
$J \approx  -\epsilon \, \beta \, \alpha^{\prime} \sin \phi \cos \phi \sin
\delta
\sim 6 \times 10^{-5} \sin \delta$.
\vskip 0.5in

\noindent
{\Large\bf IV.~Low Energy Phenomenology}
\vskip 0.1in
\noindent
In this section we study some of the consequences of Topcolor
dynamics
in low energy processes. Potentially large flavor changing neutral
currents (FCNC) arise when the quark
fields are rotated from their weak eigenbasis to their mass
eigenbasis.
In the case of Model~I, the presence of a residual $U(1)$
interacting
strongly with the third generation implies that the $Z'$ will also
couple
to leptons in order to cancel anomalies. This generates contributions
to a number of semileptonic processes. On the
other hand, in Model~II the induced
four--fermion interactions remain nonleptonic.
In all cases where quark field rotations are involved we must choose
an ansatz for the mass matrices or, equivalently, for $U_{L,R}$ and
$D_{L,R}$
as defined in previous sections.

A possible choice is to take
the square root of the CKM matrix as an indication of the order
of magnitude of the effects. This is compatible with the
matrices derived in (\ref{ul}) and (\ref{dl}), regarding the mixing
of the
third and the second generations. However, this simple ansatz
does not make a distinction between the left- and the right-handed
couplings. Such a distinction might arise in some realizations of
the models, as seen in Section~II.(C) where triangular textures were
derived. In those cases, the vanishing of some of the off--diagonal
elements precludes contributions to particle--antiparticle
mixing from nonleptonic four--fermion interactions,
although they do not have the same effect in
semileptonic transitions, where  left and right mixing factors
enter additively rather than multiplicatively. We first present the
constraints
on Topcolor FCNC from the existing data on $b\to s\g$ as well
as $B^0-\bar B^0$ and $D^0-\bar D^0$ mixing. Then we
show several predictions for semileptonic processes.

%%%%%%%%%%%%%%%%%%%%%%%%%%%%%%%%%%%%%%%%%%%%%%%%%%%%%%%%%%%%%%%%%%%%%
\vskip 0.3in
\noindent
{\large\em A.~Constraints from $b\to s\g$}
\vskip 0.1in
The recent measurement of the inclusive branching ratio for the
process
$b\to s\g$ \cite{cleo_bsg}, puts severe constraints on a variety of
extensions of the SM \cite{vs_bsgamma}. We study here the constraints
on
Topcolor models. In doing so we will neglect possible long distance
contributions to the $b\to s\g$ rate.
 Long distance effects have been estimated in the
literature \cite{long_dist} to be somewhere between $5\%$ to $50\%$
of the rate and would loosen these constraints.

The effective hamiltonian for $b\to s\g$ transitions is given by
\cite{Grinstein,CFMRS,BMMP}
\bz
H_{\rm eff}=-\fr{4}{\sqrt{2}}G_F
V_{ts}^{*}V_{tb}\sum_{i}C_i(\m)O_i(\m)
\label{heff0}
\ez
At the weak scale, the only contributing  operator is
\bz
O_7=\fr{e}{16\p^2} m_b\left(\bar{s}\s_{\m\n}b_R\right) F^{\m\n}.
\label{o7}
\ez
However, when evolving down to the low energy scale $\m\simeq m_b$,
$O_7$
will mix with the gluonic penguin operator
\bz
O_8=\fr{g}{16\p^2} m_b\left(\bar
s_\a\s_{\m\n}T^{a}_{\a\b}b_{R\b}\right)
G^{\m\n}_{a}. \label{o8}
\ez
as well as with the four-quark operators $O_1-O_6$
\cite{Grinstein,CFMRS,BMMP}.

The Topcolor~I contributions to $C_7(M_W)$ have been previously
considered
in \cite{TopC1,Kominis}. They arise from the couplings of $\phi_1$
and
$\phi_2$ to quarks in (\ref{e1}). Their charged components, the
top-pion $\tilde{\pi}^+$ and
the charged scalar $\tilde{H}^+$,
give additional penguin diagrams where they
replace the $W$. They also generate contributions to the new
operators
$O'_7$ and $O'_8$ that are obtained by switching the chirality of
quarks
in (\ref{o7}) and (\ref{o8}).
If one neglects the running from the top-pion
and $\tilde{H}^+$ mass scales down to $M_W$, the coefficients of $O_7$ and
$O'_7$
now take the form \cite{Kominis}
\by
C_7(M_W)&=&-\fr{1}{2}A(x_W)+\fr{D_{L}^{bs*}}{V^*_{ts}}
\left(\fr{v_w}{f_{\tilde{\p}}}\right)^2
\left[\fr{m^{*}_{b}}{m_b}\left(B(x_{\ctp})-B(x_{\ch2})\right)-
\fr{1}{6}A(x_{\ctp})\right]
\label{c7_tc}\\
C'_7(M_W)&=&-\fr{D_{R}^{bs*}}{V^*_{ts}}\left(\fr{v_w}{f_{\tilde{\p}}}
\right)^2
\left[\fr{1}{6}
%% FOLLOWING LINE CANNOT BE BROKEN BEFORE 80 CHAR
A(x_{\ch2})-\fr{m^{*}_{b}}{m_b}\left(B(x_{\ctp})-B(x_{\ch2})\right)
\right]\label{c7p_tc}
\ey
where $x_i=m_{t}^{2}/m_{i}^{2}$
and the functions $A(x)$ and  $B(x)$ are given in \cite{Grinstein}.
Here $D_{L}$ and $D_R$ are the matrices defining the rotation from
the
weak to the mass eigenbasis in the down sector, defined in a way
analogous to eq.(\ref{udiag}) for the up sector,
and $f_{\tilde{\pi}}\sim 50$~GeV is the top-pion decay constant.
The parameter $m_b^*$ is proportional to the couplings
of $\phi_1$ to $b_R$ and $\phi_2$ to $t_R$ which are only induced by instantons
and mixing effects. For definiteness, we shall assume that the $b$ quark mass
is mainly generated by the instanton dynamics, with a piece of
${\cal O}(1)$~GeV coming from the explicit Higgs Yukawa coupling. These
two effects add with an unknown sign, so, under our assumptions,
the ratio $m_b^*/m_b$ lies in the range $0.8 \simlt m_b^*/m_b \simlt 1.2$. We
have used the value 0.8 in our numerical estimates. Variation
of $m_b^*/m_b$ in the above range tends to tighten the constraints we
report, but not substantially.

In order to account for the renormalization group evolution of $C_7$
and $C'_7$
down to the low energy scale we need to know also the
coefficients of $O_8$ and $O'_8$, including Topcolor contributions.
At the $M_W$-scale these
coefficients, $C_8(M_W)$ and $C'_8(M_W)$, can be obtained from
$C_7(M_W)$ and $C'_7(M_W)$ by simply replacing $A\to D$ and $B\to E$,
where the functions $D(x)$ and $E(x)$ are also defined in
\cite{Grinstein}.
At the scale $\m\simeq m_b$, $O_7$ mixes with $O_8$ as well as with
the
four-quark operators. The complete leading logarithmic approximation,
within the SM first obtained in \cite{CFMRS},
gives
\bz
C_7(m_b)=\h^{16/23}C_7(M_W) +
\fr{8}{3}\left(\h^{14/23}-\h^{16/23}\right)C_8(M_W) +\sum_{i=1}^{8}
h_i\h^{p_i}
\label{c7mb}
\ez
where $\eta=\alpha_s(M_W)/\alpha_s(m_b)$.
The coefficients $h_i$ and $p_i$ can be found in \cite{BMMP}.
The ``wrong" chirality operator $O'_7$ mixes exclusively with $O'_8$,
giving
\bz
C'_7(m_b)=\h^{16/23}C'_7(M_W) +
\fr{8}{3}\left(\h^{14/23}-\h^{16/23}\right)C'_8(M_W)\label{c7pmb}
\ez

The current experimental information on the inclusive $b\to s\g$ rate
comes
from the recent CLEO measurement \cite{cleo_bsg},
$ Br(b\to s\g)=(2.32\pm 0.57\pm 0.35)\times 10^{-4}$
which can be translated into $95\%$ confidence level upper and lower
limits
as
\bz
1\times 10^{-4} < Br(b\to s\g) < 4.2\times 10^{-4} \label{limits}
\ez

Normalized by the inclusive semileptonic branching ratio, the $b\to
s\g$
branching fraction can be written as
\bz
\fr{Br(b\to s\g)}{Br(b\to ce\n)}=\fr{|V^*_{ts}V_{tb}|^2}{|V_{cb}|^2}
\fr{6\a_{em}}{\p g(z)} \left( |C_7(m_b)|^2 + |C'_7(m_b)|^2\right)
\label{ratio}
\ez
where $g(z)=1-8z^2+8z^6-z^8-24z^4\ln z$, with $z=m_c/m_b$, is a
phase-space factor arising in the semileptonic branching ratio.
In order to illustrate the constraints imposed by (\ref{limits}) on
the parameters of topcolor models, we plot the allowed region in the
$D_{L}^{bs*}/V^*_{ts}-D_{R}^{bs*}/V^*_{ts}$
plane for fixed values of the charged top-pion and
charged scalar masses, $m_{\ctp}$ and $m_{\ch2}$.  The top-pion mass
arises through the couplings of the top quark
to the Higgs boson, which are proportional to $\e$
and constitute an explicit breaking of chiral symmetry. Estimates of
this mass in the fermion loop approximation and consistent with
(\ref{e16a}) give \cite{TopC1}
$m_{\ctp}\approx (180-250)$~GeV. On the other hand, $m_{\ch2}$ can be
estimated using (\ref{e16b}). The main contribution to it comes from the
topcolor interactions.
%The expression (\ref{e16b}) is obtained in the large $N_c$
%limit and cutting off the fermion loops at $\L$. This gives only a
%very crude estimate of this mass.
For instance, in \cite{Kominis} it was shown that near
criticality and in this approximation it could be as small as
$m_{\ch2}\approx
350$~GeV. We show the constraints from $b\to s\g$ for this value as
well as for
$m_{\ch2}=1$~TeV in Fig.~2, where $D_{L}^{bs*}/V^*_{ts}$ is assumed
to be real.
The data is more constraining for larger  values
of $m_{\ch2}$. This is due to a partial cancellation of the
Topcolor effects in (\ref{c7_tc}) and (\ref{c7p_tc}), which is more
efficient for lighter scalar masses.
The second term in (\ref{c7_tc}) reduces the value of $C_7(M_W)$ with
respect to the SM for an
important range of values of $m_{\ch2}$. This can be appreciated in
Fig.~3,
where the rate is plotted versus $m_{\ch2}$ for different values of
$m_{\ctp}$
and for $D_{L}^{bs*}/V^*_{ts}=1/2$ and $D_{R}^{bs}=0$.
Nonzero values of $D_{R}^{bs}$
would compensate the cancellation bringing the rate back up in better
agreement
with experiment, given that $C'_7$ always contributes positively to
the rate.

Although a large region is still allowed
for any given pair of $\tilde{\pi}^+$ and $\tilde{H}^+$
%scalar and pseudo-scalar
masses, the data is already
constraining $D_{L}^{bs*}/V^*_{ts}$ to be positive or small.
A simple ansatz for the mixing matrices $D_L$ and $D_R$ is to assume
they are
of the order of the square root of the CKM mixing martix. This gives
$|D_{L,R}^{bs}|\simeq (1/2) |V_{ts}|$. As a reference, this is
indicated by the
square in Fig.~2. Triangular textures as the ones discussed in
Section~II.(C),
allow one of the mixing factors to be very small. The $b\to s\g$ data
then
implies a constraint on the other factor that can be extracted from
Fig.~2. These textures seem to be necessary to accomodate
$B^0-\bar{B}^0$
mixing, as we will see next.

\vskip 0.3in
\noindent
{\large\em B.~Constraints from Nonleptonic Processes}
\vskip 0.1in
At low energies the Topcolor interactions will induce four-quark
operators
leading to nonleptonic processes.
In order to study the phenomenology of these interactions it is
useful to
divide them in three categories: $\bar{D}D\,\bar{D}D$,
$\bar{U}U\,\bar{D}D$ and
$\bar{U}U\,\bar{U}U$.  Rotation to the mass eigenbasis will give new
four-quark
operators  now involving the second and first families. Although
suppressed by
mixing factors, these operators can in principle give contributions
to low
energy processes.
In Topcolor~I, the down-down operators give rise to potentially large
corrections to $B_d$ and $B_s$ mixing. They also
induce transitions not present in the SM to lowest order in $G_F$,
most notably $b\to
ss\bar{d}$, although with very small branching ratios.
In the second category, the up-down operators give vertices that
induce
corrections to processes allowed at tree level in the SM. These corrections
are of ${\cal O}(10^{-3})$ relative to the SM amplitudes,
and are therefore very hard to observe,
given that they appear in channels like $b\to
c\bar{c}s$. The effects of the operators in
these two categories are not sizeable in
Topcolor~II, since the right-handed down quarks do not couple to the
strong $SU(3)$ and there is no strong $U(1)$.
Finally, in the up-up operators, the most interesting process is
$D^0-\bar{D}^0$ mixing given that it is extremely suppressed in the
SM.

\vskip0.3in
\noindent
{\bf 1. $B^0-\bar{B}^0$ Mixing}
\vskip0.1in
\noindent
The most important effects of Topcolor in $B^0-\bar{B}^0$ mixing are
due
to the scalar sector generated at low energies. They were studied in
detail in Ref.~\cite{Kominis}. The field $\phi_2$
in (\ref{e1}) contains  the two neutral $\bar{b} b$ bound states
$\tilde{H}^0, A^0$,
which in the approximation (\ref{scal_spect0}) can be rather light.
When the quarks are rotated to their mass eigenbasis, flavor changing
couplings
of $\tilde{H}^0, A^0$ are generated. They induce a contribution to the
$B^0-\bar{B}^0$
mass difference given by \cite{Kominis}
\bz
\fr{\D m_B}{m_B}=\fr{7}{12}\fr{m_{t}^2}{f_{\pi}^2m_{\tilde H^0}^2}\;
\d_{bd}B_Bf_{B}^2
\label{d_mb}
\ez
where\footnote{The numerical coefficient on the right-hand-side
of the expression (\ref{d_mb}) inadvertently
appears as 5/12 in \cite{Kominis}, rather than the
correct 7/12.} $\d_{bd}\simeq |D_{L}^{bd}D_{R}^{bd}|$.
Using the experimental
measurement of $\D m_B$ \cite{PDG} one obtains the bound
\bz
\fr{\d_{bd}}{m_{\tilde{H}^0}^2}<  10^{-12}{\rm~GeV}^{-2}
\label{dmb_bound}
\ez
Thus, if $m_{\tilde{H}^0}$ is of the order of a few hundred GeV,
then (\ref{dmb_bound}) represents an important constraint on the mixing
factors.
For instance, if one naively uses the ansatz that takes the square
root of the
CKM mixing matrix for {\em both} $D_{L}$ and $D_{R}$, the bound
(\ref{dmb_bound}) is violated by one to two orders of magnitudes.
However, the triangular textures
motivated in Section~II.(C) provide a natural suppression of the
effect by
producing approximately
diagonal $D_L$ {\em or} $D_R$ matrices.  This gives $\d_{bd} \approx 0$
and avoids the bound altogether.

\vskip0.2in
\noindent
{\bf 2. $D^0-\bar{D}^0$ Mixing}
\vskip0.1in
\noindent
At the charm quark mass scale the dominant effect in flavor changing
neutral
currents is due to the flavor changing couplings of top-pions. In the
case of
Topcolor~I the operator inducing  $D^0-\bar{D}^0$ mixing can be
written as
\bz
{\cal H}_{eff}=\frac{1}{2m^2_{\tilde\pi^0}}
\fr{m^2_t}{2 f^2_{\tilde{\pi}}}\,\d_{cu}\,\bar{u}\g_5c\,\bar{u}\g_5c
\label{cu_coup}
\ez
where $\d_{cu}=(U^{tu*}_L U^{tc}_R)^2$ is the factor arising from the
rotation to the mass eigenstates. Here we have for simplicity assumed
$U^{tu*}_L U^{tc}_R=U^{tu*}_R U^{tc}_L$, since we are only interested
in the order of magnitude of the effect.
In the vacuum insertion approximation,
\bz
\langle\bar{D}^0\,|\bar{u}\g_5c\,\bar{u}\g_5c|\, D^0\rangle
=-2 f_{D}^{2}\, m_{D}^{2} \label{mael_op}
\ez
where $f_D$ is the $D$ meson decay constant.
Then the contribution of (\ref{cu_coup}) to the mass difference takes
the form
\bz
\D m_{D}^{TCI}=\fr{1}{2}\, f_{D}^{2}
m_D\,\fr{m_{t}^{2}}{f_{\tilde{\p}}^{2}
m_{\tilde{\p}^0}^{2}}\, |\d_{cu}| \label{dd_mix}
\ez
In the $\sqrt{\rm CKM}$ ansatz and for a top-pion mass of
$m_{\tilde{\p}^0}=200$~GeV we obtain
\bz
\D m_{D}^{TCI}\approx 2\times 10^{-14} {\rm~GeV} \label{tcdm}
\ez
which is approximately a factor of five  below the current
experimental limit
of $1.3\times 10^{-13}{\rm~GeV}$ \cite{PDG}.
On the other hand, the SM predicts $\D m_{D}^{SM}<10^{-15} {\rm~GeV}$
\cite{ddbar_sm}. This puts
potentially large Topcolor effects in the discovery window of future
high statistics charm experiments \cite{kaplan}.

The effect is even stronger in Topcolor~II. In this case the strong
coupling
of  the right-handed top with the left-handed charm quark induces
scalar and pseudoscalar top-pion couplings of the form
\bz
{\cal L}=
\fr{m_t}{\sqrt{2}f_{\tilde{\pi}}} \bar{c}_L (\pi_s^0 + i \pi_p^0)
t_R \,+ {\rm  h.c.}
\label{cu_coup2}
\ez
The operator contributing to $\D m_D$ is now
\by
%{\cal H}_{eff}=&-&\frac{1}{2m^2_{\tilde\pi^0}}
%\fr{m^2_t}{2 f^2_{\tilde{\pi}}} \left(
%U^{cu*}_L U_{R}^{tc}U_{L}^{cu*}U_{R}^{tc}\;\bar{u}_L c_R\bar{u}_Lc_R
%+(U_{R}^{tu*})^2\;\bar{u}_Rc_L\bar{u}_Rc_L \right. \\ \nonumber
%& &+ \left. 2 U_{L}^{cu*}U_{R}^{tc}U_{R}^{tu*}\;\bar{u}_R
%c_L\bar{u}_Lc_R\right)
{\cal H}_{eff}& = & -\frac{m_t^2}{f_{\tilde{\pi}}^2 m_{\tilde{\pi}}^2} U_L^{cc}
U_R^{tu*} U_L^{cu*} U_R^{tc} \bar{u}_L c_R \bar{u}_R c_L
\ey
Using the $\sqrt{\rm CKM}$ ansatz, one observes that in this
case the $D$ meson mass difference is typically larger by a factor
$1/\lambda^4$ compared to the Topcolor I scenario (with the
Wolfenstein parameter $\lambda=0.22$). Thus we estimate
\bz
\D m_{D}^{TCII}\approx\D m_{D}^{TCI}\cdot\frac{1}{\lambda^4}
\sim 10^{-11}~{\rm GeV}
\ez
which violates the current experimental upper limit by about two
orders of
magnitude. This is the single most constraining piece of
phenomenology on
the Topcolor II model.
However, once again, these constraint can be avoided in models with
triangular textures in the up sector.

\vskip 0.4in
\noindent
{\large\em C.~Semileptonic Processes}
\vskip 0.1in
\noindent We study  here the  FCNC at tree level induced by the
exchange
of the $Z'$ arising in Topcolor I models.
The corresponding effective Lagarangian is given in (\ref{lzbeff}).
After rotation to the mass eigenstates, (\ref{lzbeff}) generates
four-fermion
interactions leading to FCNC. With the exception of the effect on the
$\U(1S)$ leptonic branching ratio, the cases considered in what
follows are of
this type.
\vskip 0.1in

\noindent
{\em\bf 1.~ $\U(1S)\to \ell^+\ell^-$}
\vskip 0.1in
%In the following subsection we shall briefly discuss the leptonic
%decays of the $\U(1S)$.
Although these processes do not involve
FCNC, they still receive additional contributions from $Z'$
exchange in Topcolor I type models.
The resulting modification of the $\t^+\t^-$ rate
violates lepton universality. Experimental results
on $\U(1S)\to \ell^+\ell^-$ might therefore yield important
constraints
on model parameters, which are independent of quark mixing factors.

The $\U(1S)\to \ell^+\ell^-$ amplitude can in general be written as
\begin{equation}\label{ayll}
{\cal A}(\U(1S)\to \ell^+\ell^-)=-\frac{4\pi\alpha}{3M^2_\U}
\langle 0|(\bar bb)_V|\U(\epsilon)\rangle
\left[ r_V(\bar\ell\ell)_V+r_A(\bar\ell\ell)_A \right]
\end{equation}
Note that the axial vector piece of the b-quark current does not
contribute.
For the dominant photon exchange contribution $r_V=1$ and $r_A=0$.
In the case of the $\tau$-lepton mode these couplings are modified
by the effective $Z'$ interaction in (\ref{lzbeff}) into
\begin{equation}\label{rvra}
r_V=1-\frac{3\kappa_1}{16\alpha}\frac{M^2_\U}{M^2_{Z'}}\qquad
r_A=-\frac{\kappa_1}{16\alpha}\frac{M^2_\U}{M^2_{Z'}}
\end{equation}
The $Z'$ contribution to the electron and muon modes are suppressed
by a factor of $\tan^2\theta'$, which we neglect.

Using (\ref{ayll}) and
\bz
\langle 0|(\bar bb)^\mu_V|\U(\epsilon)\rangle = i F_{\U}\;
M_{\U}\;\e^\m
\label{up_dc}
\ez
one finds for the decay rate
\begin{equation}\label{gyll}
\Gamma(\U(1S)\to
\ell^+\ell^-)=\frac{4\pi\alpha^2}{27}\frac{F^2_\U}{M_\U}
\sqrt{1-4\frac{m^2}{M^2_\U}}\left[r^2_V\left(1+2\frac{m^2}{M^2_\U}
\right)+
r^2_A\left(1-4\frac{m^2}{M^2_\U}\right)\right]
\end{equation}
where $m$ is the lepton mass.
{}From (\ref{rvra}) and (\ref{gyll}) we see that the leading $Z'$
effect
is given by the interference of the $Z'$-exchange with the photon
amplitude. This interference is destructive, reducing the $\tau$ rate
in comparison with the electron and muon modes according to
\begin{equation}\label{rtmth}
\frac{\Gamma(\U(1S)\to\tau^+\tau^-)}{\Gamma(\U(1S)\to\mu^+\mu^-)}=
\sqrt{1-4\frac{m^2_\tau}{M^2_\U}}\left(1+2\frac{m^2_\tau}{M^2_\U}
\right)
\left[1-\frac{3}{8}\frac{\kappa_1}{\alpha}\frac{M^2_\U}{M^2_{Z'}}
\right]
\end{equation}
In the Standard Model this ratio is slightly reduced by a phase space
factor which amounts to 0.992. The lepton universality violating
$Z'$ effect leads to an additional suppression. For $\kappa_1=1$
and $M_{Z'}=500$~GeV this suppression is $\sim 2\%$.
Experimentally, the ratio is measured to be \cite{PDG,cleo_uptau}
\begin{equation}\label{rtmex}
\frac{\Gamma(\U(1S)\to\tau^+\tau^-)}{\Gamma(\U(1S)\to\mu^+\mu^-)}=
1.05\pm 0.07
\end{equation}
Presently, the error is still too large for a useful constraint
to be derived from this measurement. However, a sensitivity at the
percent level, which would start to become binding, does not seem
to be completely out of reach. It is interesting to note that the
central value in (\ref{rtmex}) actually exceeds unity. If this
feature
should persist as the error bar is reduced, it would tend to
sharpen the constraint on the negative $Z'$ effect discussed above.
Generally speaking, it seems quite plausible that a decay like
$\U(1S)\to\tau^+\tau^-$ could potentially yield important
constraints on new strong dynamics associated with the third
generation.
We have illustrated such a possibility within the framework of the
Topcolor~I scenario. In any case, a more precise measurement of
the ratio in (\ref{rtmex}) could give useful information on this type
of physics and is  very desirable.
\vskip 0.1in

\noindent
 {\em\bf 2.~ $B_s\to \ell^+\ell^-$}
\vskip 0.1in
The part of the effective interaction given in (\ref{lzbeff}) that
is inducing $B_s\to\tau^+\tau^-$ can be written as
\begin{equation}\label{hbstau}
{\cal L}_{eff,Z'}=\frac{\pi\kappa_1}{12 M^2_{Z'}}\left(
D^{bb\ast}_L D^{bs}_L(\bar bs)_{V-A}-2 D^{bb\ast}_R D^{bs}_R
(\bar
bs)_{V+A}\right)\left((\bar\tau\tau)_{V-A}+2(\bar\tau\tau)_{V+A}\right
)
\end{equation}
after performing the rotation to mass eigenstates. Using the fact
that
only the axial vector part of the quark current contributes to the
hadronic matrix element and
\bz
\langle 0|\bar{b}\g_\m\g_5 s|B_s(P_\m)\rangle =i f_{B_s} P_\m
\label{decons}
\ez
the Topcolor contribution to the amplitude for $B_s\to\tau^+\tau^-$
is
\begin{equation}\label{abstau}
{\cal A}^{TC}(B_s\to\tau^+\tau^-) =
\frac{\pi\kappa_1}{12 M^2_{Z'}} \delta_{bs} f_{B_s} P_\mu
\left((\bar\tau\tau)_{V-A}+2(\bar\tau\tau)_{V+A} \right)
\end{equation}
where we defined
\bz
\delta_{bs}=D_{L}^{bb\ast}D_{L}^{bs}+2D_{R}^{bb\ast}D_{R}^{bs}
\label{delta}
\ez
Assuming that only Topcolor contributes, the width is given by
\begin{equation}\label{gbstau}
\Gamma(B_s\to\tau^+\tau^-)=\frac{\pi\kappa^2_1}{288 M^4_{Z'}}
f^2_{B_s}
m_{B_s} m^2_\tau |\delta_{bs}|^2\sqrt{1-4\frac{m^2_\tau}{m^2_{B_s}}}
\end{equation}
Using $f_{B_s}=0.23$~GeV one gets
\bz
\Gamma(B_s\to \tau^+\tau^-)= 7\times 10^{-3}
\; |\delta_{bs}|^{2}\;
\frac{\kappa_{1}^{2}}{M_{Z'}^{4}} \;{\rm GeV}^5 \label{gamma_tt}
\ez
For the neutral mixing factors we make use of the CKM square root
ansatz ($\sqrt{\rm CKM})$.
This choice is rather general in this case given that (\ref{delta})
involves
a sum of left and right contributions and
 will not vanish when the textures are triangular as in
(\ref{m_trian}).
We still have the freedom of the relative sign between the elements
of  $D_{L}$
and $D_{R}$ in (\ref{delta}). This introduces an uncertainty of a
factor of $3$
in the amplitude.
Taking  $\kappa_1\simeq {\cal O}(1)$ we get
\bz
Br(B_s\to \tau^+\tau^-)\approx
\left\{
\begin{array}{c}
 1\;(0.1)\times 10^{-3}\;\;\; {\rm for }\;\; M_{Z'}=\; 500 \,{\rm GeV}
\\
\; 6\;(0.7)\times 10^{-5}  \;\;\; {\rm for }\;\; M_{Z'}=1000 \,{\rm
GeV}
\end{array}
\right.
\ez
where we have used the positive (negative) relative sign in
(\ref{delta}).
The SM prediction is $BR^{\rm SM}\approx 10^{-6}$ \cite{BBL,AGM}.

On the other hand, first and second generation leptons
couple to the weaker $U(1)_{Y2}$.
The corresponding amplitudes are similar to (\ref{abstau}) with
$\tau$
replaced by $e$ or $\mu$, except that they carry an additional
factor of $(-\tan^2\theta')$.
For instance, for $B_s\to \mu^+\mu^-$ one obtains
\bz
Br(B_s\to \mu^+\mu^-)=\left(\frac{m_{\mu}}{m_{\tau}}\right)^2
\left(1-4\frac{m^2_\tau}{m^2_{B_s}}\right)^{-1/2} \;\tan^4\q '
\times Br(B_s\to \tau^+\tau^-) \label{br_mm}
\ez
The choice $\kappa_1=1$ corresponds to $\tan\q '\approx 0.1$. This
gives a
suppression factor of $\approx 5\times 10^{-7}$ with respect to the
$\tau$
mode. Of this suppression, a factor of $3.5\times 10^{-3}$ comes from
helicity.
This is not present in the $b\to s \ell^+\ell^-$ decays, which makes
the $\mu$
modes more accessible. The SM predicts
$BR^{\rm SM}(B_s\to \mu^+\mu^-)\approx 4\times 10^{-9}$
\cite{BBL,AGM}.

Finally, the rates for the $B_d$ purely leptonic
modes are obtained by replacing $D_{L,R}^{bs}$ by
$D_{L,R}^{bd}$. In the $\sqrt{\rm CKM}$ ansatz this represents a
suppression of
$\approx 10^{-2}$ in the branching fractions with respect to the
$B_s$ case.
\vskip 0.3in

\noindent
{\em\bf 3.~ $B\to X_{s}\; \ell^+\ell^-$}
\vskip 0.1in
Using the normalization of the effective hamiltonian as in
(\ref{heff0}),
the operators contributing to these processes are $O_7$, as given by
eq.(\ref{o7}), and
\by
O_9&=&\fr{e^2}{16\p^2} \left(\bar{s}\g_\m
b_L\right)\left(\bar{\ell}\g^\m\ell\right) \\
O_{10}&=&\fr{e^2}{16\p^2}\left(\bar{s}\g_\m
b_L\right)\left(\bar{\ell}\g^\m\g_5\ell\right)\label{opbasis} \nn
\ey
The contact interaction induced by the $Z'$ exchange gives new
contributions to
the coefficient functions $C_9$ and $C_{10}$ at $\mu=M_W$ as well as
non-zero
values for the coefficients of the new operators \cite{gb_bsll}
\by
O'_9&=&\fr{e^2}{16\p^2} \left(\bar{s}\g_\m
b_R\right)\left(\bar{\ell}\g^\m\ell\right) \label{w_basis_2}\\
O'_{10}&=&\fr{e^2}{16\p^2}\left(\bar{s}\g_\m
b_R\right)\left(\bar{\ell}\g^\m\g_5\ell\right) \label{w_basis_3}
\ey
These are given by
\by
C_{9}^{TC}(M_W)&=&\frac{1}{2}\frac{D_{L}^{bs*}D_{L}^{bb}}{V_{ts}^{*}V_
{tb}} \;
F \\
C_{10}^{TC}(M_W)&=&\frac{1}{6}\frac{D_{L}^{bs*}D_{L}^{bb}}{V_{ts}^{*}V
_{tb}} \;
F \\
C_{9}^{'TC}(M_W)&=&-1\frac{D_{R}^{bs*}D_{R}^{bb}}{V_{ts}^{*}V_{tb}}\;
F
\\
C_{10}^{'TC}(M_W)&=&-\frac{1}{3}\frac{D_{R}^{bs*}D_{R}^{bb}}{V_{ts}^{*
}V_{tb}}\
F
\ey
where we defined
\bz\label{fdef}
F=\frac{4\pi^2\; v_{w}^2}{\alpha}\;\frac{\kappa_1}{M^2_{Z'}}
\ez
Here $v^2_w=(2\sqrt{2} G_F)^{-1}=(174~GeV)^2$.
(Note that in this section it is assumed that
an adequate
order-of-magnitude estimate of the novel effects can be obtained by only
considering the consequences of
$Z^{\prime}$ exchange.
In particular, the scalar boundstate contributions
to the coefficient functions $C_7$ and $C_7^{\prime}$
(cf. eq.(\ref{c7_tc}), (\ref{c7p_tc})) are not taken into account.)
Eq. (\ref{fdef}) applies to the case of the $\tau$-lepton. For $e$ or
$\mu$, $F$ carries an additional factor of $(-\tan^2\theta')\approx
-0.01$.
The dilepton mass distribution has the form
\by
\frac{d Br(b\to
s\ell^+\ell^-)}{ds}&=&K(1-s)^2\sqrt{1-\frac{4x}{s}}\left\{\left(
|C_9|^2+|C'_9|^2-|C_{10}|^2-|C'_{10}|^2\right)6x\right. \nn\\
&+&\left(|C_9|^2+|C'_9|^2+|C_{10}|^2+|C'_{10}|^2\right)\left[(s-4x)+
\left(
1+\frac{2x}{s}\right)(1+s)\right] \nn \\
&+& \left. 12C_7{\rm
Re}[C_9]\left(1+\frac{2x}{s}\right)+\frac{4|C_7|^2}{s}\left(1+\frac{2x}
{s}\right)(2+s)
%\right\} \label{dil_dist}
\right. \label{dil_dist}
\ey
where $s=q^2/m_{b}^{2}$ and $x=m_{\ell}^{2}/m_{b}^{2}$. The factor
$K$ is given
by
\bz\label{fac_k}
K=\frac{\alpha^2}{4\pi^2}\left|\frac{V^*_{ts}V_{tb}}{V_{cb}}\right|^2
\frac{Br(b\to ce\nu)}{g(z)}
\ez
where the function $g(z)$, $z=m_c/m_b$, can be found after eq.
(\ref{ratio}).
The SM contributions to $b\to s\ell^+\ell^-$ reside in the
coefficients $C_7$, $C_9$ and $C_{10}$. For the present discussion we
will neglect QCD effects. This gives a reasonable approximation,
which
is completely sufficient for our purposes. We use here the
coefficient
functions from \cite{BM} in the limit of vanishing $\alpha_s$.
To illustrate the possible size of the new physics effect we choose
again
the $\sqrt{\rm CKM}$ ansatz. In this case we also have to choose the
sign of
$D_{L}^{bs}$, which is taken to be positive.
Furthermore we set $\kappa_1=1$.
To estimate the order of magnitude of the branching ratios, we simply
integrate the dilepton invariant mass spectrum in (\ref{dil_dist})
from $4x$ to $1$.
Numerical results are given
in Table~I. The branching ratios are similar to those for  $B_s\to
\tau^+\tau^-$, given that the partial helicity suppression is
balanced by the
phase space suppression in the three body decay. There are no
presently
published limits on any of the $\tau$ channels.
On the other hand, the angular information in these decays provides a
sensitive
test of the chirality of the operators involved. This is true not
only for
the inclusive decays \cite{ali}
but also for exclusive modes like $B\to K^*\ell^+\ell^-$,
where the SM lepton asymmetry is very distinct in a region where
hadronic
matrix elements can be reliably predicted \cite{gb_bsll,liu}.

\begin{table}
\centering
\begin{tabular}{|c|c|c|}
\hline
$M_{Z'}[{\rm GeV}]$&$Br(b\to s\tau^+\tau^-)$&$Br(b\to s\mu^+\mu^-)$
\\ \hline
$500$&$1.4\times 10^{-3}$&$6.7\times 10^{-6}$\\
$1000$ &$0.9\times 10^{-4}$&$6.0\times 10^{-6}$\\
SM &$3.7\times 10^{-7}$&$6.3\times 10^{-6}$\\
\hline
\end{tabular}
\vskip 0.75truecm
\caption{Estimates of inclusive branching ratios for $b\to
s\ell^+\ell^-$ in
the SM and Topcolor. }
\end{table}

\vskip 0.1in
\noindent
{\em\bf 4.~ $B\to X_s \nu\bar{\nu}$}
\vskip 0.1in
The decay $b\to s\nu\bar\nu$ could have an important contribution
from the
$\tau$ neutrino which couples
strongly to the $U(1)_1$ in Topcolor I models. The Topcolor amplitude
can be
derived from (\ref{lzbeff}) and reads
\bz
{\cal A}^{TC}(b\to s\nu_\tau\bar{\nu}_\tau)=-\frac{\pi\kappa_1}{12
M^2_{Z'}}
\left(g_v\;(\bar sb)_V+g_a\;(\bar
sb)_A\right)\;(\bar\nu_\tau\nu_\tau)_{V-A}
\label{bsnunu}
\ez
where
\by
g_v&=&D_{L}^{bb}D_{L}^{bs*}-2D_{R}^{bb}D_{R}^{bs*}
\nn\\
g_a&=&-\left(D_{L}^{bb}D_{L}^{bs*}+2D_{R}^{bb}D_{R}^{bs*}\right) \nn
\ey
On the other hand the SM amplitude is
\bz
{\cal A}^{SM}(b\to s\nu_\tau\bar{\nu}_\tau)=
\frac{G_F}{\sqrt{2}}\frac{\alpha}{2\pi\sin^2\theta_W}
V^\ast_{ts}V_{tb}
X(x_t) (\bar sb)_{V-A} (\bar\nu_\tau\nu_\tau)_{V-A} \label{bsnnsm}
\ez
where $x_t=m^2_t/M^2_W$ and the Inami-Lim function $X(x)$ is
given by
\begin{equation}\label{ilfxx}
X(x)=\frac{x}{8}\left[-\frac{2+x}{1-x}+\frac{3x-6}{(1-x)^2}\ln
x\right]
\end{equation}
Taking the mixing factors to be
\bz
\d_{bs}= D_{L}^{bb}D_{L}^{bs*}=D_{R}^{bb}D_{R}^{bs*}\sim
\frac{1}{2}|V_{ts}|\sim \frac{1}{2}|V_{cb}|
\label{nn_dbs}
\ez
we have
\begin{equation}\label{atsbsnn}
\left|\frac{{\cal A}_{TC}}{{\cal A}_{SM}}\right|=
\frac{2\pi^2\sin^2\theta_W v^2_w}{3\alpha
X(x_t)}\frac{\kappa_1}{M^2_{Z'}}
\frac{\sqrt{|g_v|^2+|g_a|^2}}{\sqrt{2}|V^\ast_{ts}V_{tb}|}
\approx 4\times 10^6 \,{\rm GeV}^2 \frac{\kappa_1}{M^2_{Z'}}
\end{equation}
The square of this ratio divided by the number of neutrinos gives an
estimate
of the ratio of branching ratios. We obtain
\bz
\frac{BR^{TC}(b\to s\n_\t\bar{\n_\t})}{BR^{SM}(b\to s\n\bar{\n})}\sim
\left\{
\begin{array}{c}
93 \kappa_{1}^{2}  \;\;\; {\rm for }\;\; M_{Z'}=\; 500 \,{\rm GeV} \\
\;\; 6 \kappa_{1}^{2}  \;\;\; {\rm for }\;\; M_{Z'}=1000 \,{\rm GeV}
\end{array}
\right.
\ez
Estimates of this mode in the SM  give $BR^{SM}\simeq 4.5\times
10^{-5}$
\cite{BBL,AGM}.
\vskip 0.1in

\noindent
{\em\bf 5.~ $K^+\to \pi^+ \nu\bar{\nu}$}
\vskip 0.1in
As in $B\to X_s \nu\bar{\nu}$, we are concerned with the contact term
involving
$\tau$ neutrinos, given that they constitute the most important
Topcolor
contribution. The Topcolor amplitude is given by
\bz
{\cal A}^{TC}(K^+\to \pi^+ \nu_\t\bar{\nu}_\t) =
-\frac{\pi\kappa_1}{12 M_{Z'}^{2}} \; \d_{ds} \;
\langle\pi^+(k)|(\bar sd)_V|K^+(p)\rangle
\;\;(\bar\nu_\tau\nu_\tau)_{V-A} \label{atc_nunu}
\ez
where now $\d_{ds}=D_{L}^{bs*}D_{L}^{bd}-2D_{R}^{bs*}D_{R}^{bd}$. The
hadronic matrix element in (\ref{atc_nunu}) can be written in terms
of the
one entering the semileptonic decay
\bz
\langle\pi^+(k)|(\bar sd)^\mu_V|K^+(p)\rangle
=\sqrt{2}\langle\pi^0(k)|(\bar su)^\mu_V|K^+(p)\rangle
=f_+(q^2)(p+k)^\mu
\label{hme}
\ez
and the form-factor $f_+(q^2)$ is experimentally well known. In any
case it
will cancel when taking the ratio to the SM amplitude. For one
neutrino species
this is given by
\bz
{\cal A}^{SM}(K^+\to\pi^+\nu_\tau\bar{\nu}_\tau)=
\frac{G_F}{\sqrt{2}}\frac{\alpha}{2\pi\sin^2\theta_W}
\sum_{j=c,t} V^\ast_{js}V_{jd} X(x_j)
\langle\pi^+|(\bar sd)_V|K^+\rangle (\bar\nu_\tau\nu_\tau)_{V-A}
\label{kpnnsm}
\ez
where $x_j=m_{j}^{2}/M^{2}_{W}$ and $X(x)$ is the Inami-Lim function
defined in (\ref{ilfxx}). Here we have neglected QCD corrections and
the
$\tau$-lepton mass effects. Since
only the vector quark current contributes to the exclusive
transition, the
Dirac structure in the Topcolor and SM amplitudes is the same.
The ratio of the Topcolor amplitude to the SM is then
\bz
\left. \frac{{\cal A}_{TC}}{{\cal A}_{SM}}\right|_{\nu_\tau}=
-\frac{2\pi^2\sin^2\theta_W v^2_w}{3\alpha}\frac{\kappa_1}{M^2_{Z'}}
\frac{\delta_{ds}}{S}
\ez
where we defined
\bz
S=\sum_{j=c,t}V^{*}_{js}V_{jd} X(x_j) \label{s_def}
\ez
For $m_t=175$~GeV, we have $|S|\approx 10^{-3}$
and the ratio can be expressed as
\bz
\left|\frac{{\cal A}^{TC}}{{\cal A}^{SM}}\right|_{\nu_\tau}=
6\times 10^9\; |\d_{ds}|\;
\frac{\kappa_1}{M_{Z'}^{2}}{\rm~GeV}^2 \label{ratio_2}
\ez
The $\sqrt{CKM}$ ansatz yields
\bz
\d_{ds} =-\frac{1}{4}\lambda^5\;(\frac{3}{4}\lambda^5) \label{mix}
\ez
when choosing positive (negative) relative signs between the two
terms entering
in $\d_{ds}$.
This gives, for $M_{Z'}=500$~GeV and  $\kappa_1=1$,
a ratio of amplitudes which is about $3\; (9)$.
\\
In the SM one expects $Br(K^+\to\pi^+\nu\bar\nu)\sim 10^{-10}$ \cite{BLO}.
Presently, the experimental upper limit is $5.2\times 10^{-9}$ \cite{PDG}.
Experiments are under way at Brookhaven National Laboratory (BNL)
to reach the sensitivity necessary to observe $K^+\to\pi^+\nu\bar\nu$
if it occurs at the SM level \cite{LV}. Any significant deviation from
the SM expectation should then also show up in these experiments.

\vskip 0.7in
\noindent
{\Large\bf V.~Conclusions}
\vskip 0.1in
\noindent
In this paper we have given a treatment of the low energy
phenomenological implications of Topcolor models, with possible
generic implications of extended Technicolor schemes.  We have also given an
effective Lagrangian analysis of the
boundstates
in Topcolor models.  This provides a point of departure for further
studies.
For example, we have not examined the
question of whether the $\theta$ term in TopC gives rise to
novel, non--CKM observable
CP--violation.
The potentially observable effects we have considered
arise because
the current basis of quarks
and leptons of the third generation experience new strong forces.
When we diagonalize the mass matrix to arrive at the mass basis
there will be induced flavor changing interactions.
These are largely amplified by the boundstate formation,
e.g.,
effects like
$B\overline{B}$ mixing are induced by coloron exchange,
but the formation of low mass top-pions gives the dominant
contribution
in a channel contained by the coloron exhange. Hence, it really
suffices to consider the dominant effect in the context of
the effective Lagrangian.  In the semi-leptonic processes the effects
are controlled by the $Z'$ exchange, and the top-pions effects
are not dominant.
 Topcolor, to an extent, explains the suppression
of the $3\rightarrow 2,1$ mixing angles, though without further
assumptions
about the origin of generational structure it cannot distinguish
between
first and second generations.

We have also sketched how the Topcolor scheme can impose textures
upon the mass matrix which is inevitable due to the  gauge
quantum numbers that distinguish generations.  A chiral--triangular
texture  emerges as a natural possibility which
can suppress dangerous processes such as $B\overline{B}$ mixing.
Without this natural source of FCNC suppression, the model would
require
fine--tuning or an {\em ad hoc} texture assumption.

We view Topcolor as a family of models. We have discussed two
classes:
Topcolor I models which involve an additional $U(1)'$
to tilt the chiral condensate into the $\overline{t}t$
flavor direction; Topcolor~II, based upon the gauge group
$SU(3)_Q\times SU(3)_1\times SU(3)_2
\times U(1)_{Y} \times SU(2)_L $, where there is only the
conventional
$U(1)_{Y}$, and no strong additional $U(1)$.  These latter models
admit a rather intriguing anomaly
cancellation solution in which the $(c, s)_{L,R}$ doublets are
treated differently under the strong  $SU(3)_1\times SU(3)_2$
structure
and a triplet of ``Q-quarks'' occurs which can condense to break
Topcolor.
Topcolor I has a number of sensitive implications in semileptonic
processes which we have detailed; Topcolor II gives only
a handful of novel effects in nonleptonic processes, most notably in
$D^0-\bar{D}^0$ mixing.

In general, it appears that Topcolor does not produce an overwhelming
degree of obvious new physics  in the low energy spectrum.
Therefore, significant GIM violating dynamics in the third generation
does not seem to imply large observable deviations from standard model
in low energy experiments at present.
The most sensitive effects are semileptonic
and trace to the more model dependent $Z'$.
These effects may be shared by other
generational $U(1)$ models (e.g., Holdom's \cite{Holdom}).
The purely nonleptonic effects are harder to disentagle from
electroweak physics.  Thus, the electroweak scale itself and
the scale of the top quark mass are places to look for the new
physics.
Our analysis shows that dramatic new physics can emerge at
High--$p_T$
in the third generation having eluded detection in sensitive low
energy experiments.
However, in a large number of future high statistics experiments
there are potential signatures of this new physics and these should
be sought.

\vskip .5in
\noindent
{\bf Acknowledgements}
\vskip .1in
\noindent
This work was performed at the
Fermi National Accelerator Laboratory,
which is operated by Universities
Research Association, Inc., under
contract DE-AC02-76CHO3000 with
the U.S. Department of Energy. D.K. thanks the Fermilab
Theory Group for their
warm hospitality. He further acknowledges support
from NSF contract PHY-9057173
and DOE contract DE-FG02-91ER40676 (while at Boston University)
and from the
German DFG under contract number Li519/2-1.

\newpage
\vskip 0.1in
\noindent
\begin{center}
{\bf Figure Captions}
\end{center}
\vskip .5in

\noindent
Figure~1:  Phase diagram of the Topcolor Model~I discussed in
Section~II.(A).

\vskip .2in
\noindent
Figure~2: Constraints from $b\to s\g$ in the $(D_L^{bs}/V_{ts},
D_R^{bs}/V_{ts})$
plane. Here
$D_L^{bs}/V_{ts}$ is assumed to be real. The allowed region between
the two
ellipses
corresponds to the $95\%$ C.L. lower and upper bounds from
Ref.~\cite{cleo_bsg}. Figure~2(a) is for $m_{\ch2}=350$~GeV, whereas
Figure~2(b) is for $m_{\tilde{H}^0}=1000$~GeV. In both cases,
$m_{\ctp}
=180$~GeV.
\vskip .2in
\noindent
Figure~3:  The predictions for the $b\to s\g$ branching ratio in the
Topcolor Model~I as a function of $m_{\ch2}$. In this figure we have
taken
$D_L^{bs}/V_{ts}=0.5$, $D_R^{bs}/V_{ts}=0$, $m_{b}^{*}=0.8 m_b$,
$m_t=175$~GeV.
The top-pion mass is taken to be: $m_{\ctp}=180$~GeV (dashed),
$m_{\ctp}=240$~GeV (solid) and $m_{\ctp}=300$~GeV (dotted).
The horizontal lines correspond to the $95\%$ C.L. upper and lower
limits
from Ref.~\cite{cleo_bsg}.

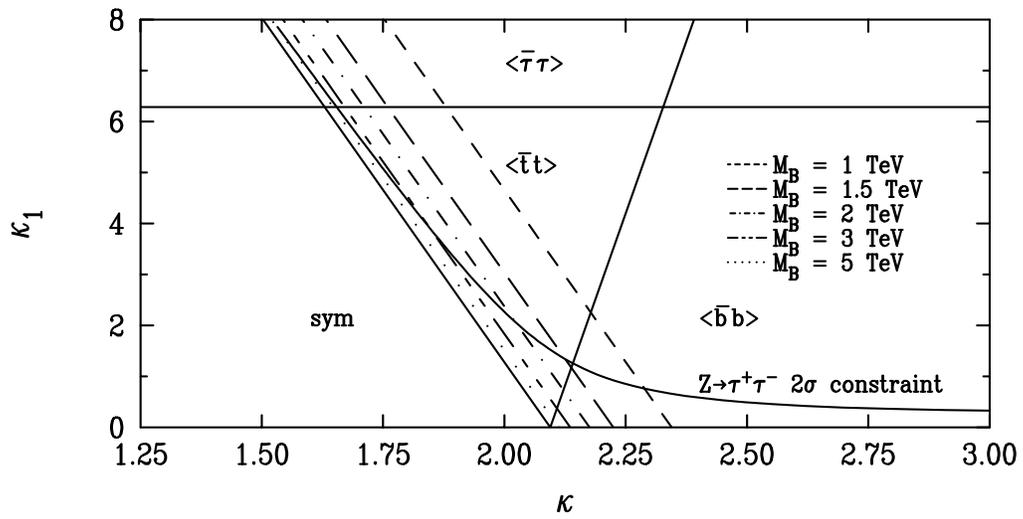
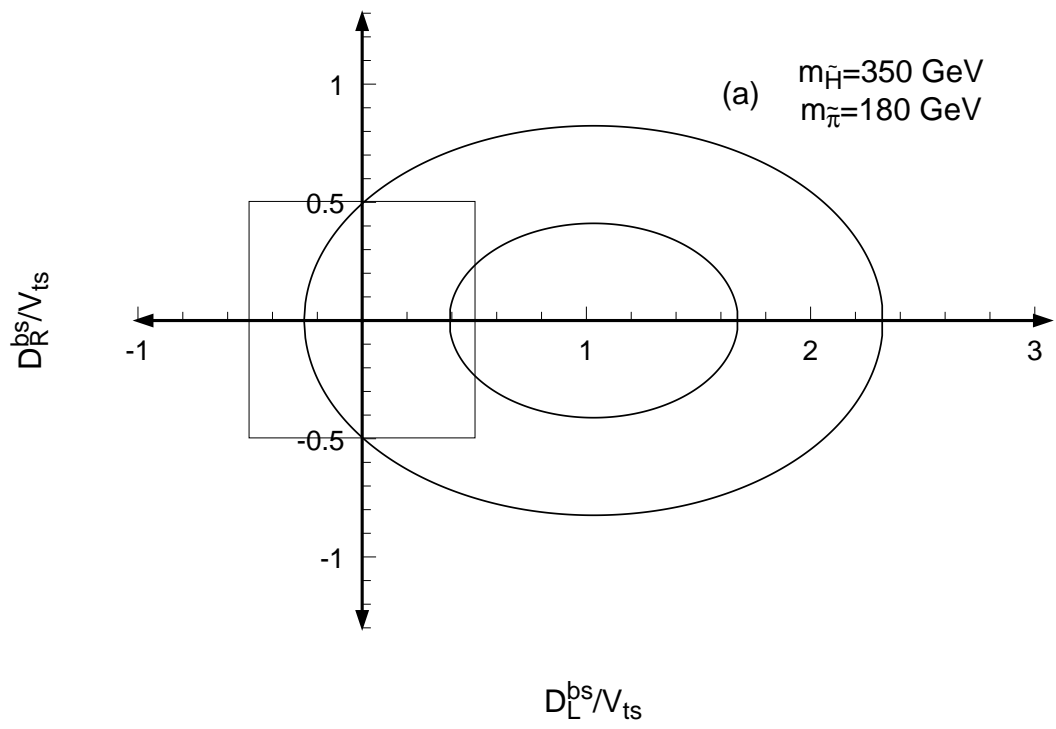
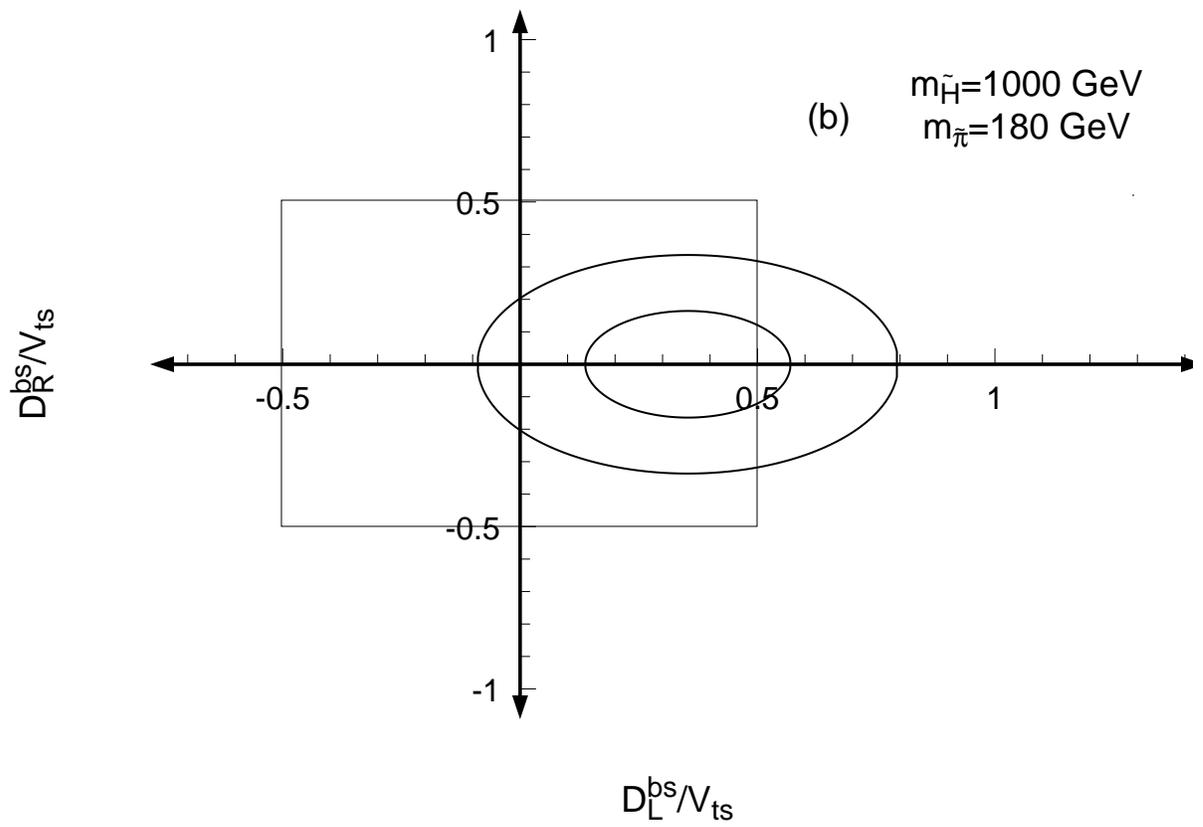
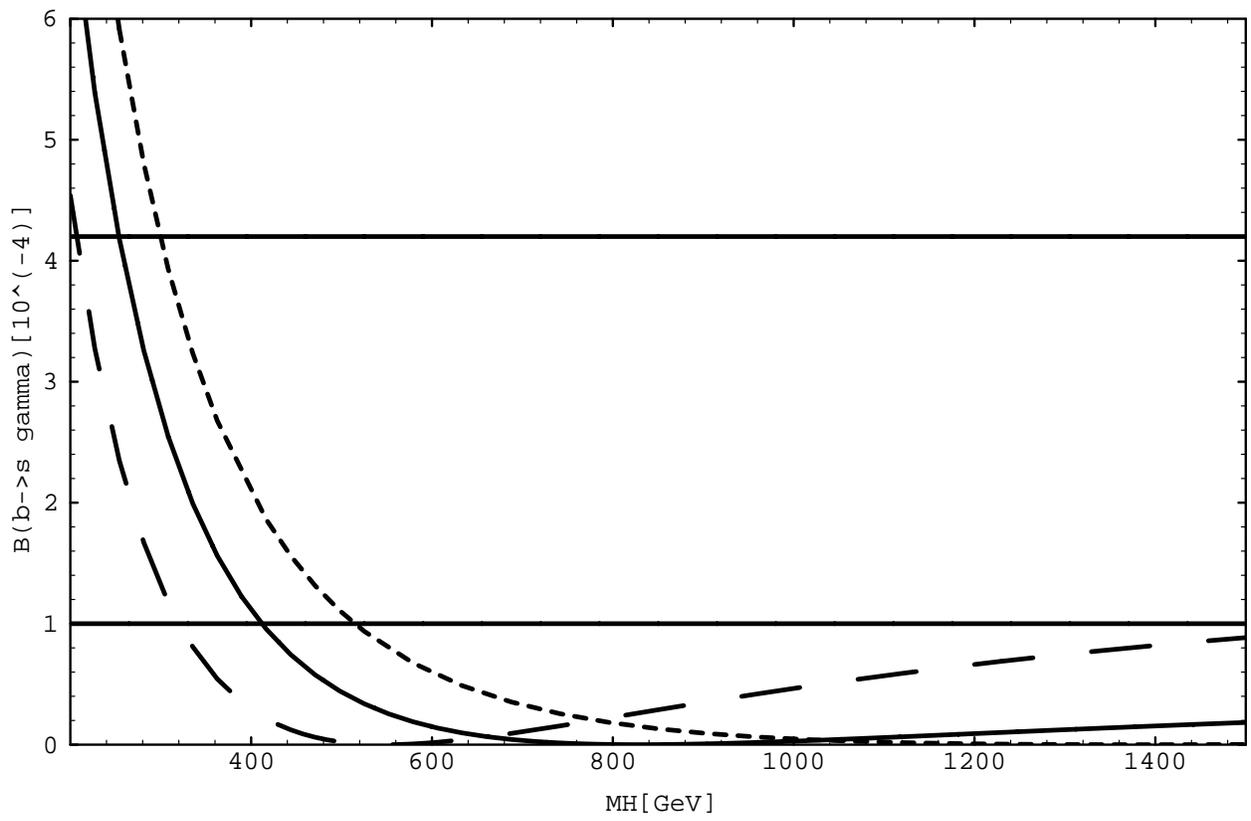
\end{document}